\documentclass[reprint, aip, apl]{revtex4-1} 
\usepackage{graphicx} 
\usepackage{multirow}
\usepackage{amsmath} 
\usepackage{soul} 
\usepackage{verbatim} 
\usepackage{bm}
\usepackage{siunitx} 
\usepackage[space]{grffile}
\usepackage[colorlinks=true, linkcolor=blue, citecolor=gray]{hyperref}
\usepackage[capitalize]{cleveref} 
\usepackage{xr} 
\usepackage{color}
\usepackage{tikz} 
\usetikzlibrary{calc, positioning} 
\graphicspath{{../Figures2/}}

\newcommand{\ket}[1]{\left\lvert #1 \right\rangle}

\newcommand{\MHz}{\mathrm{MHz}} 
\newcommand{\GHz}{\mathrm{GHz}} %Time
 
\newcommand{\us}{\mu\mathrm{s}}
\newcommand{\ns}{\mathrm{ns}}

\newcommand{\Gsps}{\mathrm{GSample/s}}

\newcommand{\dBm}{\mathrm{dBm}} 
\newcommand{\dB}{\mathrm{dB}}

\newcommand{\fpump}{f_\mathrm{pump}} 
\newcommand{\Ppump}{P_\mathrm{pump}}
\newcommand{\GTWPA}{G_\mathrm{JTWPA}} 
\newcommand{\etapre}{\eta_\mathrm{pre}}
\newcommand{\etaTWPAd}{\eta_\mathrm{JTWPAd}}
\newcommand{\etapost}{\eta_\mathrm{post}}

\newcommand{\epsdzero}{\varepsilon_\mathrm{d0}}
\newcommand{\epsdone}{\varepsilon_\mathrm{d1}}
\newcommand{\phidzero}{\phi_\mathrm{d0}}
\newcommand{\phidone}{\phi_\mathrm{d1}} 
\newcommand{\epszero}{\varepsilon_0}
\newcommand{\eps}{\varepsilon}

\newcommand{\tauup}{\tau_\mathrm{up}}
\newcommand{\taud}{\tau_\mathrm{d}} 
\newcommand{\taucost}{\tau_\mathrm{c}}
\newcommand{\tauint}{T}%\tau_\mathrm{int}}
\newcommand{\taubuffer}{\tau_\mathrm{buffer}}

\newcommand{\rhozeroone}{\rho_{01}}
\newcommand{\sigmaz}{\sigma_\mathrm{z}} 
\newcommand{\sigmam}{\sigma_\mathrm{m}}
\newcommand{\SNR}{\mathrm{SNR}} 
\newcommand{\etae}{\eta_\mathrm{e}}
 
\newcommand{\Vint}{V_\mathrm{int}}
\newcommand{\Vintzero}{V_{\mathrm{int}, \ket{0}}}
\newcommand{\Vintone}{V_{\mathrm{int}, \ket{1}}}

\newcommand{\gammam}{\beta_\mathrm{m}} 
\newcommand{\alphag}{\alpha_{\ket{0}}}
\newcommand{\alphae}{\alpha_{\ket{1}}}

\begin{document} 
\title{General method for extracting the quantum efficiency
of dispersive qubit readout in circuit QED}

\author{C.~C.~Bultink}
\author{B.~Tarasinski}
\affiliation{QuTech, Delft University of Technology, P.O. Box 5046, 2600 GA Delft, The Netherlands}
\affiliation{Kavli Institute of Nanoscience, Delft University of Technology, P.O. Box 5046, 2600 GA Delft, The Netherlands}
\author{N.~Haandb{\ae}k}
\affiliation{Zurich Instruments AG, Technoparkstrasse 1, 8005 Z\"urich, Switzerland}

\author{S.~Poletto}
\affiliation{QuTech, Delft University of Technology, P.O. Box 5046, 2600 GA Delft, The Netherlands}
\affiliation{Kavli Institute of Nanoscience, Delft University of Technology, P.O. Box 5046, 2600 GA Delft, The Netherlands}

\author{N.~Haider}
\affiliation{QuTech, Delft University of Technology, P.O. Box 5046, 2600 GA Delft, The Netherlands}
\affiliation{Netherlands Organisation for Applied Scientific Research (TNO),
P.O. Box 155, 2600 AD Delft, The Netherlands}

\author{D.~J.~Michalak}
\affiliation{Components Research, Intel Corporation, 2501 NW 229th Ave, Hillsboro, OR 97124, USA}
\author{A.~Bruno}

\author{L.~DiCarlo}
\affiliation{QuTech, Delft University of Technology, P.O. Box 5046, 2600 GA Delft, The Netherlands}
\affiliation{Kavli Institute of Nanoscience, Delft University of Technology, P.O. Box 5046, 2600 GA Delft, The Netherlands}

\date{today}

\begin{abstract} We present and demonstrate a general three-step
method for extracting the quantum efficiency of dispersive qubit readout in
circuit QED.  We use active depletion of post-measurement photons and optimal
integration weight functions on two quadratures to maximize the signal-to-noise ratio of non-steady-state homodyne measurement.  We derive analytically and demonstrate
experimentally that the method robustly extracts the quantum efficiency for
arbitrary readout conditions in the linear regime.  We use the proven method to
optimally bias a Josephson traveling-wave parametric amplifier and to quantify the
different noise contributions in the readout amplification chain. 
\end{abstract}

\maketitle

\begin{figure*}[ht!]    
\centering      
\includegraphics{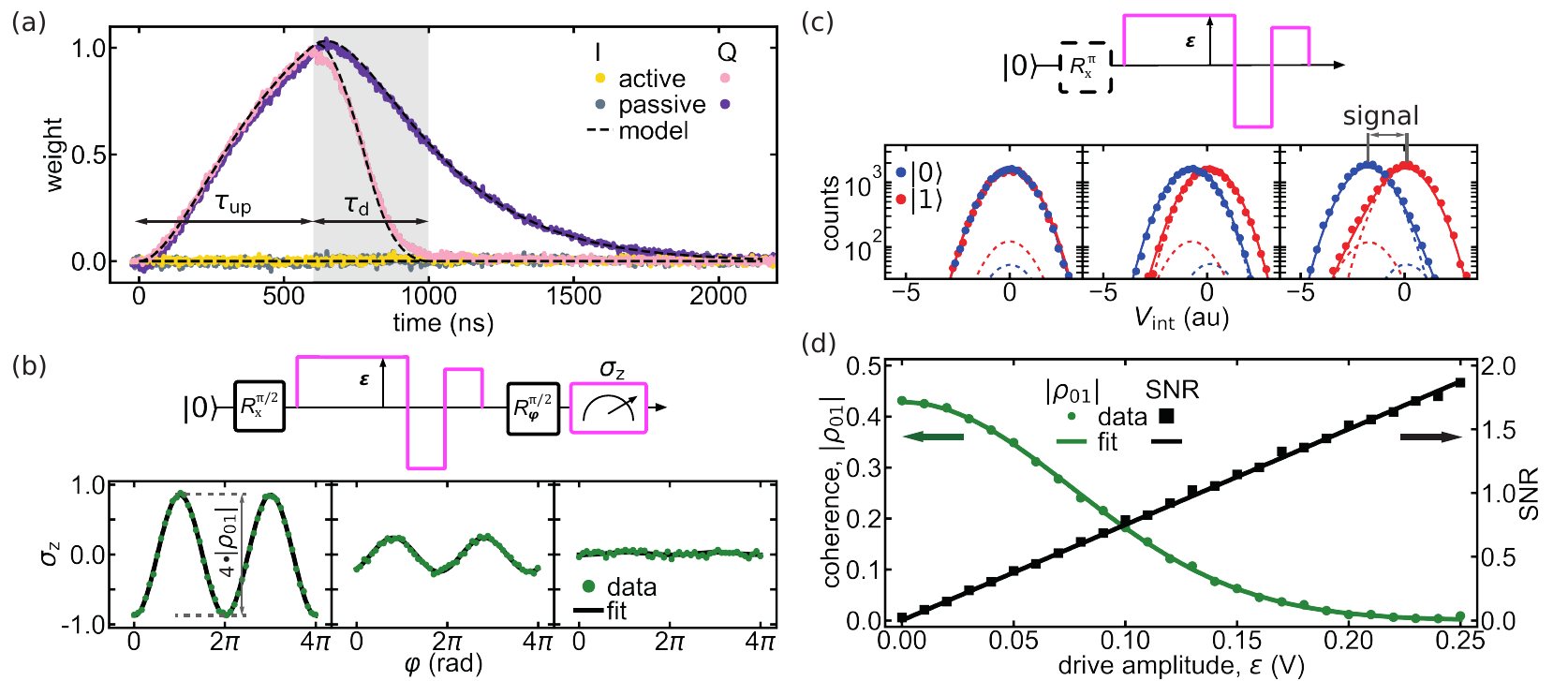}
\caption{
\label{fig:method}    
	The three-step method for extracting the quantum
	efficiency with active photon depletion. 
	(a) Calibration of the optimal weight functions for the in-phase quadrature I and out-of-phase quadrature Q for active depletion (passive depletion is shown for reference). 
	The measurement pulse consists of a ramp-up of duration $\tauup=600~\ns$ and two $200~\ns$ depletion segments
	($\taud=400~\ns$). 
	The weight functions show the dynamics of the information gain during readout and the
	effect of the active photon depletion (grey area). 
	Dashed black curves are extracted from a linear model (see supplementary material).
	(b) Study of dephasing under variable-strength weak measurement.
	Observed Ramsey fringes at from left to right $\eps=0.0,0.12,0.25$ V.   
	The measurement pulse, globally scaled with $\eps$, is embedded in a fixed-length
	($\tauint=1100~\ns$) Ramsey sequence with final strong fixed-amplitude
	measurement. 
	The azimuthal angle $\varphi$ of the final $\pi/2$ rotation is
	swept from 0 to $4\pi$ to discern deterministic phase shifts and dephasing.
	The coherence $\left|\rhozeroone\right|$ is extracted by fitting each fringe with
	the form
	$\sigmaz=2\left|\rhozeroone\right|\cos\left(\varphi+\varphi_0\right)$. 
	(c) Study of signal-to-noise ratio of variable-strength weak measurement. 
	Histograms of $2^{15}$ shots at from left to right: $\eps=0.0,0.12,0.25$ V. 
	The qubit is prepared in $\ket{0}$ without (blue) and in $\ket{1}$ with a $\pi$ pulse (red). 
	Each measurement record is integrated in real time with the weight functions of (a) during
	$\tauint=1100~\ns$ to obtain $\Vint$. Each histogram (markers) is fitted with the sum of
	two Gaussian functions (solid lines), whose individual components are indicated by the dashed lines. From the fits we get the signal, distance between the main Gaussian for $\ket{0}$ and $\ket{1}$, and
	noise, their average standard deviations. 
	(d) Quantum efficiency extraction.
	Coherence data is fitted with the form $\left|\rhozeroone\right|=b
	\mathrm{e}^{-\eps^2/2\sigma^2}$ and signal-to-noise data with the form
	$\SNR=a\eps$. From the best fits we extract $\etae = a^2 
	\sigma^2/2=0.165 \pm 0.002$.} 
\end{figure*}

Many protocols in quantum information processing, like quantum error correction~\cite{Divincenzo09,Terhal15}, require rapid interleaving of qubit gates and measurements. 
These measurements are ideally nondemolition, fast, and high fidelity.
In circuit QED~\cite{Blais04,Wallraff04,Koch07}, a leading platform for quantum computing, nondemolition readout is routinely achieved by off-resonantly coupling a qubit to a resonator. 
The qubit-state-dependent dispersive shift of the resonator frequency is inferred by measuring the resonator response to an interrogating pulse using homodyne detection.
A key element setting the speed and fidelity of dispersive readout is the quantum efficiency~\cite{Clerk10}, which quantifies how the signal-to-noise ratio is degraded with respect to the limit imposed by quantum vacuum fluctuations. 

In recent years, the use of superconducting parametric amplifiers~\cite{Castellanos-Beltran08,Vijay09,Bergeal10,Mutus14,Eichler14} as the front end of the readout amplification chain has boosted the quantum efficiency towards unity, leading to readout infidelity on the order of one percent~\cite{Johnson12,Riste12} in individual qubits.
Most recently,  the development of traveling-wave parametric amplifiers~\cite{Macklin15,Vissers16} (TWPAs) has extended the amplification bandwidth from tens of MHz to several GHz and with sufficient dynamic range to readout tens of qubits. 
For characterization and optimization of the amplification chain, the ability to robustly determine the quantum efficiency is an important benchmark.

A common method for quantifying the quantum efficiency $\eta$ that does not rely on calibrated noise sources compares the information obtained in a weak qubit measurement (expressed by the signal-to-noise-ratio SNR) to the dephasing of the qubit (expressed by the decay of the off-diagonal elements of the qubit density matrix)~\cite{Gambetta06,FootnoteBultink17}, $\eta=\frac{\SNR^2}{4\gammam}$, with $\mathrm{e}^{-\gammam} =\frac{\left|\rhozeroone(T)\right|}{\left|\rhozeroone(0)\right|}$, where $T$ is the measurement duration. 
Previous experimental work~\cite{Vijay11,Hatridge13,Jeffrey14,Macklin15} has been restricted to fast resonators driven under specific symmetry conditions such that information is contained in only one quadrature of the output field and in steady state.
To allow in-situ calibration of $\eta$ in multi-qubit devices under development~\cite{Liu17, Reagor17, Takita17, Neill17,Versluis17}, a method is desirable that does not rely on either of these conditions.

In this Letter, we present and demonstrate a general three-step method for extracting the quantum efficiency of linear dispersive readout in cQED.
Our method disposes with previous requirements in both the dynamics and the phase space trajectory of the resonator field, while requiring two easily met conditions:
the depletion of resonator photons post measurement~\cite{McClure16,Bultink16}, and the ability to perform weighted integration of both quadratures of the output field~\cite{Ryan15,Magesan15}.
We experimentally test the method on a qubit-resonator pair with a Josephson TWPA (JTWPA)~\cite{Macklin15} at the front end of the amplification chain. 
To prove the generality of the method, we extract a consistent value of $\eta$ for different readout drive frequencies and drive envelopes. 
Finally, we use the method to optimally bias the JTWPA and to quantify the different noise contributions in the readout amplification chain.

We first derive the method, obtaining experimental boundary conditions.
For a measurement in the linear dispersive regime of cQED, the internal field $\alpha(t)$ of the readout resonator, driven by a pulse with envelope $\eps f(t)$ and detuned by $\Delta$ from the resonator center frequency, is described by~\cite{Gambetta06,FriskKockum12}
\begin{equation}
\frac{\partial\alpha_{\ket{0}/\ket{1}}}{\partial t} = -i \eps
f(t)-i(\Delta\pm\chi)\alpha(t)-\frac{\kappa}{2}\alpha(t), \label{eq:cavity-eom}
\end{equation}
where $\kappa$ is the resonator linewidth and $2\chi$ is the dispersive shift. 
The upper (lower) sign has to be chosen for the qubit in the ground $\ket{0}$
[excited $\ket{1}$] state.  
We study the evolution of the SNR and the measurement-induced dephasing as a function of the drive amplitude $\eps$, while keeping $T$ constant. We find that the SNR scales linearly, $\SNR=a\eps$, and that coherence elements exhibit a Gaussian dependence, $\left|\rhozeroone(T,\eps)\right|=
\left|\rhozeroone(T,0)\right| \mathrm{e}^{-\frac{\eps^2}{2\sigmam^2}}$, with $a$ and $\sigmam$ dependent on $\kappa$, $\chi$, $\Delta$, and $f(t)$.
Furthermore, we find (Supplementary material)
\begin{equation} 
	\eta=\frac{\SNR^2}{4\gammam}=\frac{\sigmam^2a^2}{2}
	\label{eq:main-eta-formula}
\end{equation} 
provided two conditions are met. The conditions are: i) optimal integration functions~\cite{Ryan15,Magesan15} are used to optimally extract information from both quadratures, and ii) the intra-resonator field vanishes at the beginning and end; i.e.,\ photons are depleted from the resonator post-measurement.

To meet these conditions, we introduce a three-step experimental method. 
First, tuneup active photon depletion (or depletion by waiting) and calibration of the optimal integration weights. 
Second, obtain the measurement-induced dephasing of variable-strength weak measurement by including the pulse within a Ramsey sequence. 
Third, measure the SNR of variable-strength weak measurement from single-shot readout histograms.

We test the method on a cQED test chip containing seven transmon qubits with dedicated readout resonators, each coupled to one of two feedlines (see supplementary material). 
We present data for one qubit-resonator pair, but have verified the method with other pairs in this and other devices. 
The qubit is operated at its flux-insensitive point with a qubit frequency $f_\mathrm{q}=5.070~\GHz$, where the measured energy relaxation and echo dephasing times are $T_1=15~\us$ and $T_\mathrm{2,echo}=26~\us$, respectively. 
The resonator has a low-power fundamental at $f_\mathrm{r,\ket{0}}=7.852400~\GHz$ ($f_\mathrm{r,\ket{1}}=f_\mathrm{r,\ket{0}}+\chi/\pi=7.852295~\GHz$) for qubit in $\ket{0}$ ($\ket{1}$), with linewidth $\kappa/2\pi=1.4~\MHz$. 
The readout pulse generation and readout signal integration are performed by single-sideband mixing.
Pulse-envelope generation, demodulation and signal processing are performed by a Zurich Instruments UHFLI-QC with 2 AWG channels and 2 ADC channels running at $1.8~\Gsps$ with $14-$ and $12-$bit resolution, respectively.

In the first step, we tune up the depletion steps and calibrate the optimal integration weights. 
We use a measurement ramp-up pulse of duration $\tauup=600~\ns$, followed by a photon-depletion counter pulse~\cite{McClure16,Bultink16} consisting of two steps of $200~\ns$ duration each, for a total depletion time $\taud=400~\ns$. 
To successfully deplete without relying on symmetries that are specific to a readout frequency at the midpoint between ground and excited state resonances (i.e., $\Delta=0$), we vary 4 parameters of the depletion steps (details provided in the supplementary material).
From the averaged transients of the finally obtained measurement pulse, we extract the optimal integration weights given by~\cite{Ryan15,Magesan15} the difference between the averaged transients for $\ket{0}$ and $\ket{1}$ [Fig.~\ref{fig:method}(a)]. 
The success of the active depletion is evidenced by the nulling at the end of $\taud$. 
In this initial example, we connect to previous work by setting $\Delta=0$, leaving all measurement information in one quadrature.

We next use the tuned readout to study its measurement-induced dephasing and SNR to finally extract $\eta$.
We measure the dephasing by including the measurement-and-depletion pulse in a Ramsey sequence and varying its amplitude, $\eps$ [Figs.~\ref{fig:method}(b)]. 
By varying the azimuthal angle of the second qubit pulse, we allow distinguishing dephasing from deterministic phase shifts and extract $\left|\rhozeroone\right|$ from the amplitude of the fitted Ramsey fringes. 
The $\SNR$ at various $\eps$ is extracted from single-shot readout experiments preparing the qubit in $\ket{0}$ and $\ket{1}$ [Figs.~\ref{fig:method}(c)]. 
We use double Gaussian fits in both cases, neglecting measurement results in the spurious Gaussians to reduce corruption by imperfect state preparation and residual qubit excitation and relaxation. 
As expected, as a function of $\eps$, $\left|\rhozeroone\right|$ decreases following a Gaussian form and the $\SNR$ increases linearly [Fig.~\ref{fig:method}(d)]. 
The best fits to both dependencies give $\etae=0.165\pm 0.002$. 
Note that both dephasing and SNR measurements include ramp-up, depletion and an additional $\taubuffer=100~\ns$, making the total measurement window $\tauint=\tauup+\taud+\taubuffer=1100~\ns$. 

\begin{figure}[ht!]   
\centering     
\includegraphics{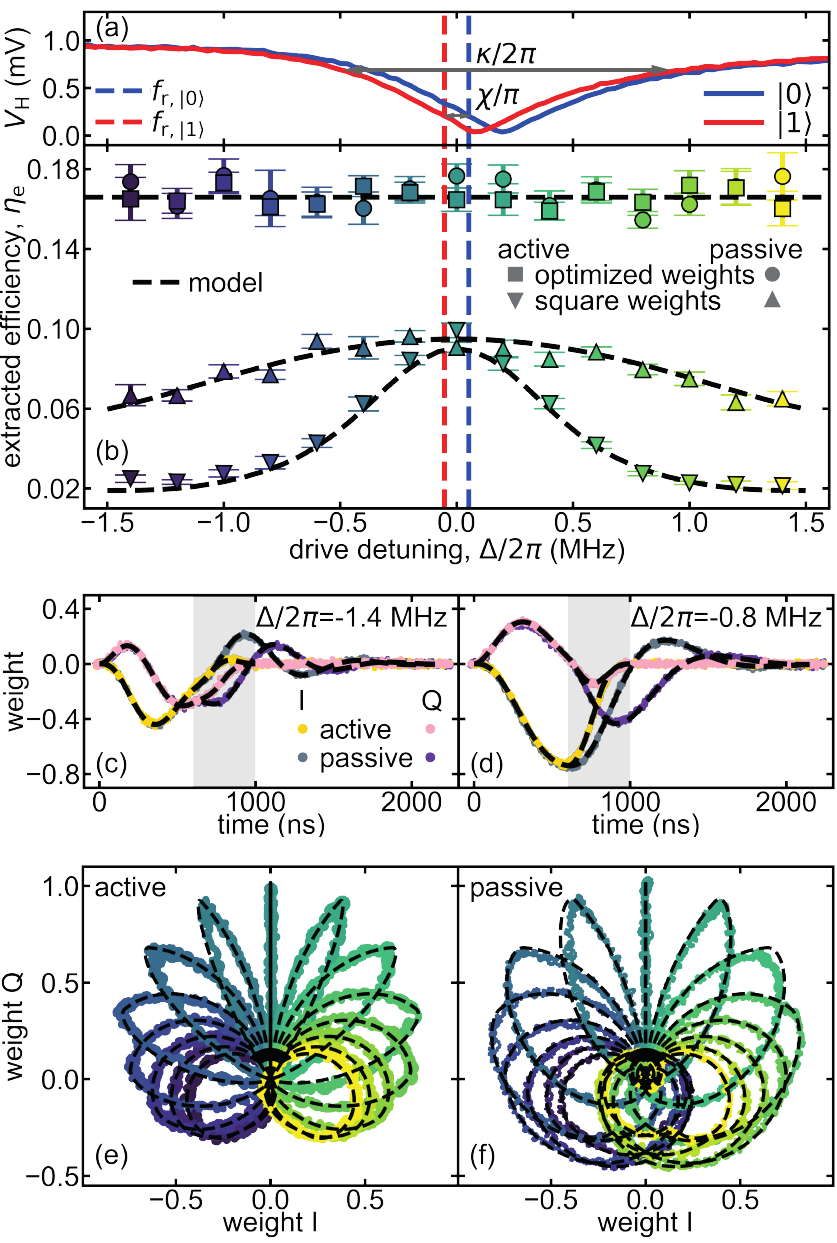}
\caption{\label{fig:detuning}   
	(a) Pulsed feedline transmission near the low-power resonator fundamentals. The qubit is prepared in $\ket{0}$ without (blue)
	and in $\ket{1}$ with a $\pi$ pulse (red). The data fits $\kappa/2\pi=1.4~\MHz$
	and $f_\mathrm{r,\ket{0}}=7.852400~\GHz$ ($f_\mathrm{r,\ket{1}}=7.852295~\GHz$),
	indicated by the dashed vertical lines. (b) Quantum efficiency extraction as a
	function of $\Delta$ using the pulse timings and three-step method of
	Fig.~\ref{fig:method}. We use both the active depletion ($\tauint=1100~\ns$) and passive depletion schemes
	($\tauint=2100~\ns$) and assess the benefit of optimal
	weights to standard square integration weights. (c,d) Optimal weight functions for I and
	Q at $\Delta/2\pi=-1.4~\MHz$, $-0.8~\MHz$ [as in Fig.~\ref{fig:method}(a)]. (e, f)
	Parametric plot of the optimal weight functions at all measured $\Delta$ [marker colors
	correspond to (a)]. Dashed black curves (b-f) are extracted from a linear model
	(see supplementary material).} 
\end{figure}

\begin{figure}[ht!]   
\centering     
\includegraphics{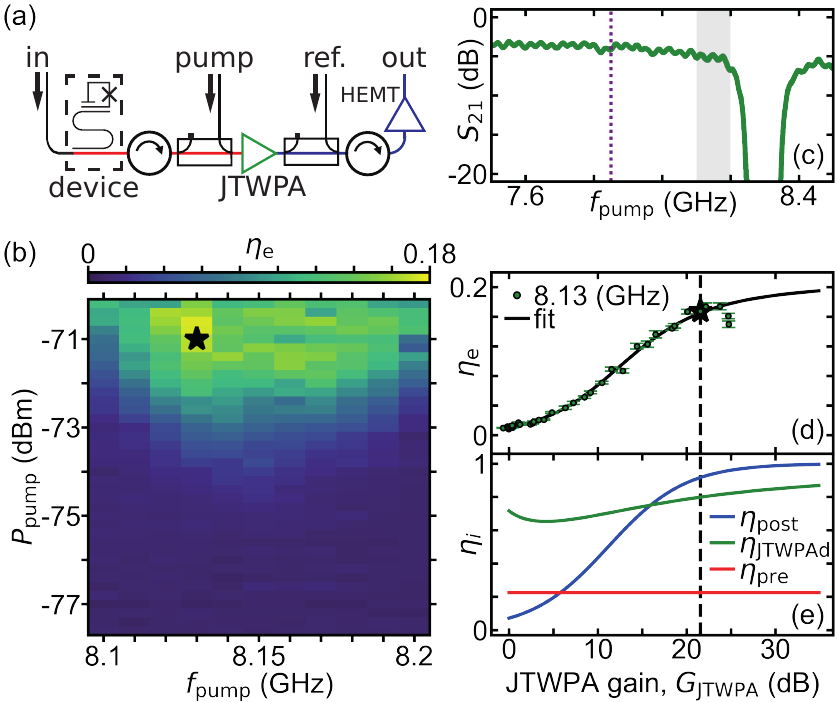}
\caption{\label{fig:TWPA}   
	JTWPA pump tuneup to maximize the quantum efficiency
	and amplification chain modeling. (a) Simplified setup diagram,  showing the
	input paths for the readout signal carrying the information on the qubit state
	and the added pump tone biasing the JTWPA. Both microwave tones are combined in
	the JTWPA amplifying the small readout signal. (b) $\etae$ as a function of pump
	power and frequency. (c) CW low-power transmission of the JTWPA showing the dip
	in transmission due to the dispersion feature near $8.3~\GHz$ and low-power
	insertion loss of $\sim4.0~\dB$ near $f_\mathrm{r,\ket{0}}$ (dashed vertical
	line). The grey area indicates the frequency range of (b). $S_{21}$ is obtained
	by measuring and comparing the output power when selecting the pump input or the
	reference input (input lines are duplicates and calibrated up to the directional
	couplers at room temperature). (d) Parametric plot of $\etae$ at
	$\fpump=8.13~\GHz$ and independently measured JTWPA gain. The fit (line) uses a
	three-stage model with
	$\eta(\GTWPA)=\etapre\times\etaTWPAd(\GTWPA)\times\etapost(\GTWPA)$ [model details
	in the main text]. (e) Plots of the best-fit $\etapre$, $\etaTWPAd(\GTWPA)$
	and $\etapost(\GTWPA)$. The stars (b, d) and vertical dashed lines (d, e)
	indicate ($\Ppump=-71.0~\dBm$, $\fpump=8.13~\GHz$, $\eta=0.1670$,
	$\GTWPA=21.6~\dB$) used throughout the experiment.
} \end{figure}

We next demonstrate the generality of the method by extracting $\eta$ as a function of the readout drive frequency.  We repeat the method at fifteen readout drive detunings over a range of $2.8~\MHz\sim\kappa/\pi\sim14 \chi/\pi$ around $\Delta=0$ [Figs.~\ref{fig:detuning}(a,b)]. 
Furthermore, we compare the effect of using optimal weight functions versus square weight functions, and the effect of using active versus passive photon depletion.  
The square weight functions correspond to a single point in phase space during $\tauint$, with unit amplitude and an optimized phase maximizing SNR. 
We satisfy the zero-photon field condition by depleting the photons actively with $\tauint=1100~\ns$ (as in Fig.~\ref{fig:method}) or passively by waiting with $\tauint=2100~\ns$. 
When information is extracted from both quadratures using optimal weight functions, we measure an average $\etae=0.167$ with $0.004$ standard deviation. 
The extracted optimal integration functions in the time domain [Figs.~\ref{fig:detuning}(c,d)] show how the resonator returns to the vacuum for both active and passive depletion. 
Square weight functions are not able to track the measurement dynamics in the time domain (even at $\Delta=0$), leading to a reduction in $\etae$. 
Figures~\ref{fig:detuning}(e,f) show the weight functions in phase space. 
The opening of the trajectories with detuning illustrates the rotating optimal measurement axis during measurement and leads to a further reduction of increase of $\etae$ when square weight functions are used. 
The dynamics and the $\etae$ dependence on $\Delta$ are excellently described by the linear model, which uses eq.~\ref{eq:cavity-eom}, the separately calibrated $\kappa$ and $\chi$ [Fig.~\ref{fig:detuning}(a)] and $\eta=0.1670$ (details in the supplementary material). 
Furthermore, the matching of the dynamics and depletion pulse parameters (see supplementary material) when using active photon depletion confirm the numerical optimization techniques. 

To further test the robustness of the method to arbitrary pulse envelopes, we have used a measurement-and-depletion pulse envelope $f(t)$ resembling a typical Dutch skyline. 
The pulse envelope outlines five canal houses, of which the first three ramp up the resonator and the latter two are used as the tunable depletion steps. Completing the three steps, we extract (see supplementary material) $\etae=0.167\pm0.005$, matching our previous value to within error.

We use the proven method to optimally bias the JTWPA and to quantify the different noise contributions in the readout chain.  
To this end, we map $\etae$ as a function of pump power and frequency, just below the dispersive feature of the JTWPA, finding the maximum $\etae=0.1670$ at ($\Ppump=-71.0~\dBm$, $\fpump=8.13~\GHz$) [Figs.~\ref{fig:TWPA}(a-c)]. 
We next compare the obtained $\etae$ at the optimal bias frequency to independent low-power measurements of the JTWPA gain $\GTWPA$ we find $\GTWPA=21.6~\dB$ at the optimal bias point. 
We fit this parametric plot with a three-stage model, with noise contributions before, in and after the JTWPA, $\eta(\GTWPA)=\etapre \times \etaTWPAd(\GTWPA) \times \etapost(\GTWPA)$. 
The parameter $\etapre$ captures losses in the device and the microwave network between the device and the JTWPA and is therefore independent of $\GTWPA$.
The JTWPA has a distributed loss along the amplifying transmission line, which is modeled as an array of interleaved sections with quantum-limited amplification and sections with attenuation adding up to the total insertion loss of the JTWPA (as in Ref.~\onlinecite{Macklin15}).
Finally, the post-JTWPA amplification chain is modeled with a fixed noise temperature, whose relative noise contribution diminishes as $\GTWPA$ is increased. 
The best fit [Figs.~\ref{fig:TWPA}(d,e)] gives $\etapre=0.22$, consistent with $50\%$ photon loss due to symmetric coupling of the resonator to the feedline input and output, an attenuation of the microwave network between device and JTWPA of $2~\dB$ and residual loss in the JTWPA of $27\%$. We fit a distributed insertion loss of the JTWPA of $4.6~\dB$, closely matching the separate calibration of $4.2~\dB$ [Fig.~\ref{fig:TWPA}(c)]. 
Finally, we fit a noise temperature of $2.6$ K, close to the HEMT amplifier's factory specification of $2.2$ K.\\

We identify room for improving $\etae$ to $\sim0.5$ by implementing Purcell filters with asymmetric coupling~\cite{Jeffrey14,Walter17} (primarily to the output line) and decreasing the insertion loss in the microwave network, by optimizing the setup for shorter and superconducting cabling between device and JTWPA.

In conclusion, we have presented and demonstrated a general three-step method for extracting the quantum efficiency of linear dispersive qubit readout in cQED.  
We have derived analytically and demonstrated experimentally that the method robustly extracts the quantum efficiency for arbitrary readout conditions in the linear regime. 
This method will be used as a tool for readout performance characterization and optimization.\\

See supplementary material for a description of the linear model, the derivation of Eq.~(2), a description of the depletion tuneup and additional figures.

\begin{acknowledgments} We thank W.~D.~Oliver for providing the JTWPA, N.~K.~Langford for experimental contributions, M.~A.~Rol for software contributions, and C.~Dickel and F.~Luthi for discussions. This research is supported by the Office of the Director of National Intelligence (ODNI), Intelligence Advanced Research Projects Activity (IARPA), via the U.S. Army Research Office Grant No. W911NF-16-1-0071. Additional funding is provided by Intel Corporation and the ERC Synergy Grant QC-lab. The views and conclusions contained herein are those of the authors and should not be interpreted as necessarily representing the official policies or endorsements, either expressed or implied, of the ODNI, IARPA, or the U.S. Government. The U.S. Government is authorized to reproduce and distribute reprints for Governmental purposes notwithstanding any copyright annotation thereon.

\end{acknowledgments}

\clearpage

\renewcommand{\thefigure}{S\arabic{figure}}
\renewcommand{\theequation}{S\arabic{equation}}
\setcounter{figure}{0}
\setcounter{equation}{0}

\section*{Supplementary material for ``General method for extracting the quantum efficiency
of dispersive qubit readout in circuit QED``}

\maketitle
This supplement provides additional sections and figures in support of claims in the main text. In Sec.~\ref{sec:basic-expr}, we present details of the linear model we use to describe the resonator and qubit dynamics during linear dispersive readout. In Sec.~\ref{sec:modelling-fig2}, we describe how we evaluated these expressions to obtain the dashed lines in Fig.~2 of the main text, to which experimental results are compared. 
In Sec.~\ref{sec:eq2}, we show that Eq.~(2) follows from the linear model. 
Sec.~\ref{sec:cost} provides the cost function used for the optimization of depletion pulses. Figure~\ref{fig:detuning_supplement} supplies the  optimized depletion pulse parameters as a function of $\Delta$ and the SNR and coherence as a function of the drive amplitude and $\Delta$. Figure~\ref{fig:amsterdam} shows the extraction of $\etae$ for an alternative pulse shape. Finally, Fig.~\ref{fig:setup} provides a full wiring diagram and a photograph of the device.

\section{Modeling of resonator dynamics and measurement signal}
\label{sec:basic-expr}

In this section, we give the expressions that model the resonator dynamics and measured signal in the linear dispersive regime.

In general, the measured homodyne signal consists of in-phase (I) and in-quadrature (Q) components, given by~\cite{FriskKockum12e}
\begin{align} 
V_{\mathrm{I}, \ket{i}}(t)&=V_0\left(\sqrt{2\kappa\eta}
  \mathrm{Re}(\alpha_{\ket{i}}(t))+n_\mathrm{I}(t)\right), \nonumber \\
V_{\mathrm{Q}, \ket{i}}(t)&=V_0\left(\sqrt{2\kappa\eta}
  \mathrm{Im}(\alpha_{\ket{i}}(t))+n_\mathrm{Q}(t)\right). \label{eq:def-viq}
\end{align} 
Here, $V_0$ is an irrelevant gain factor and $n_\mathrm{I}$, $n_\mathrm{Q}$ are continuous, independent Gaussian white noise terms with unit variance,
$\langle n_{j}(t)n_{k}(t')\rangle=\delta_{jk}\delta(t-t')$, while the internal resonator field
$\alpha_{\ket{i}}$ 
follows Eq.~(1) for $i\in \{0,1\}$. In the shunt resonator arrangement used on the device for this work, the measured signal also includes an additional term describing the directly transmitted part of the measurement pulse. We omitted this term here, as it is independent of the qubit state, and thus is irrelevant for the following, as we will exclusively encounter the signal difference $\Vintone - \Vintzero$.

For state discrimination, the homodyne signals are each multiplied with weight functions, given by the difference of the averaged signals, then
summed and integrated over the measurement window of duration $T$:
\begin{align} 
  V_{\mathrm{int}, \ket{i}}=\int_0^T w_\mathrm{I} V_{\mathrm{I}, \ket{i}} +w_\mathrm{Q} V_{\mathrm{Q}, \ket{i}} dt. \label{def-vint}
\end{align} 
The optimal weight functions~\cite{Ryan15e,Magesan15e} are given by the difference of the average signal
\begin{align}
  w_\mathrm{I/Q} = \langle V_{\mathrm{I/Q},\ket{1}}-V_{\mathrm{I/Q},\ket{0}}\rangle. \label{eq:def-wopt}
\end{align}
As an alternative to optimal weight functions, often constant weight functions are used
\begin{align}
  w_\mathrm{I} = \cos{\phi_w}, \quad w_\mathrm{Q} = \sin{\phi_w}, \label{eq:def-w-square}
\end{align}
where the demodulation phase $\phi_w$ is usually chosen as to maximize the $\SNR$ (see below).

We define the signal $S$  as the absolute separation between the average $\Vint$ for $\ket{1}$ and $\ket{0}$. In turn, we define the noise $N$ as the standard deviation of $V_{\mathrm{int, \ket{i}}}$, which is independent of $\ket{i}$. Thus,
\begin{align*}
  S& =\left|\langle \Vintone-\Vintzero \rangle\right|, \nonumber \\
  N^2  &= \langle \Vint ^ 2 \rangle -  \langle \Vint \rangle^2. 
\end{align*}
The signal-to-noise ratio $\SNR$ is then given as
\begin{align}
  \SNR = \frac{S}{N}. \label{eq:def-s-and-n}
\end{align}

The measurement pulse leads to measurement-induced dephasing. Experimentally, the dephasing can be quantified by including the measurement pulse in a Ramsey sequence. The coherence
elements of the qubit density matrix are reduced due to the pulse as~\cite{FriskKockum12e} 
\begin{align*} 
	\left|\rhozeroone(\eps)\right|=\mathrm{e}^{-\gammam} \left|\rhozeroone(\eps = 0)\right|,
\end{align*} 
where  
\begin{align}
  \gammam = 2 \chi \int_0^T \mathrm{Im}(\alphag \alphae^*) dt. \label{eq:def-gammam}
\end{align} 
Thus, $\gammam$ scales with $\eps^2$, and the coherence elements decay as a Gaussian in $\eps$.

\section{Comparison of experiment and model}
\label{sec:modelling-fig2}

We here describe how we compared the theoretical model given by the previous section and Eq.~(1) to 
the experimental data as presented in Fig.~2.

In panels (c)-(f) of Fig.~2, we compare the measured weight functions to a numerical evaluation of Eq.~(1).
The dashed lines in those panels are obtained by numerically integrating Eq.~(1), using the $\epsilon$ and $\Delta$ applied in experiment, and with the resonator parameters $\kappa$ and $\chi$ 
that are obtained from resonator spectroscopy [presented in panel (a)].
From the resulting $\alpha_{\ket{i}}$ we then evaluate Eqs.~\eqref{eq:def-viq} and \eqref{eq:def-wopt} to obtain $w_{\mathrm{I/Q}}$, 
presented in panels (c)-(f). The scale factor $V_0$ was chosen to best represent the experimental data. 

In order to model the data presented in panel (b), we further inserted the $\alpha_{\ket{i}}$ into Eqs.~\eqref{eq:def-s-and-n} and \eqref{eq:def-gammam}, and finally into
Eq.~(2) to obtain $\etae$. This step is performed for both optimal weights and constant weights, Eqs.~\eqref{eq:def-wopt}~and~\eqref{eq:def-w-square}.
As shown in Fig.~2, the result depends on pulse shape and $\Delta$ when using square weights, but does not when using optimal weights. The value for $\eta$ in Eq.~\eqref{eq:def-viq} is chosen as the average of $\etae$ for optimal weight functions, $\etae=0.167$.

\section{Derivation of equation 2}
\label{sec:eq2}

With the definitions of the previous sections, we now show that Eq.~(2) holds for arbitrary pulses and resonator parameters if optimal weight functions are used,
so that $\etae$ in Fig.~2 indeed coincides with $\eta$ in Eq.~\eqref{eq:def-viq}.

Using optimal weight functions, we can evaluate Eq.~\eqref{eq:def-s-and-n} in terms of $\alpha_{\ket{i}}$  by inserting Eqs.~\eqref{eq:def-wopt} and \eqref{eq:def-viq}, 
obtaining for the signal $S$:
\begin{align*}
  S_\mathrm{opt} = 2\kappa \eta V_0^2 \int_0^T \left| \alpha_\mathrm{\ket{1}}-\alpha_\mathrm{\ket{0}} \right| ^2 dt.
\end{align*}
For the noise $N$, we obtain
\begin{align*} 
  N^2_\mathrm{opt} & = V_0^2\Bigg\langle\int_0^T \Big(w_\mathrm{I} n_\mathrm{I} + w_\mathrm{Q} n_\mathrm{Q}\Big)^2 dt\Bigg\rangle \nonumber \\
	  &= 2 \kappa \eta V_0^4 \int_0^T \left|
	\alpha_\mathrm{\ket{1}}-\alpha_\mathrm{\ket{0}} \right| ^2 dt,
\end{align*}
where we used the white noise property of $n_{\mathrm{I, Q}}(t)$.

The SNR is then given by
\begin{align} 
  \SNR_\mathrm{opt} = \frac{S_\mathrm{opt}}{N_\mathrm{opt}} =
\sqrt{2\kappa \eta \int_0^T \left| \alpha_\mathrm{\ket{1}}-\alpha_\mathrm{\ket{0}} 
  \right| ^2 dt}. \label{eq:snr-opt}
\end{align}
Note that the $\alpha_{\ket{i}}$ scale linearly with the amplitude $\eps$ due to the linearity of Eq.~(1), so that the SNR scales linearly with $\eps$ as well.

We now show that the $\gammam$ and SNR are related by Eq.~(2), independent of resonator and pulse parameters.
For that, we need to make use of constraint (ii), namely that the resonator fields $\alpha_{\ket{i}}$
vanish at the beginning and end of the integration window. We then can write
\begin{align*}
  0 &= \left[ \left|\alphag - \alphae \right|^2 \right]_0^T \\
  &=\int_0^T \partial_t \left| \alphag - \alphae \right|^2 dt \\ 
  & = 2\int_0^T \mathrm{Re}\left( (\alphae^* - \alphag^*)\partial_t (\alphae - \alphag)\right)dt,
\end{align*}
where the first equality is ensured by requirement (ii), and the second equality follows from rewriting as the integral of a differential.

We insert the differential equation Eq.~(1) into this expression, obtaining
\begin{align*}
    &\mathrm{Re}\int_0^T \left(\alphae^* - \alphag^*\right) \times  \\
	&\Bigg(\left(-i\Delta - \frac{\kappa}{2}\right) \left(\alphae - \alphag\right) 
	- i \chi \left(\alphae + \alphag\right)\Bigg) dt = 0.
\end{align*}
Isolating the $\kappa$ term and dropping purely imaginary $\Delta$ and $\chi$ terms, we obtain
\begin{align*}
  & \frac{\kappa}{2} \int_0^T \left| \alphae - \alphag \right|^2 dt \\
  = &  - \mathrm{Re} \left( i\chi \int_0^T \left( \alphae + \alphag\right) (\alphae^* - \alphag^*) dt \right) \\
  = &  - \mathrm{Re}\left( i \chi \int_0^T \left(|\alphae|^2 - |\alphag|^2
	 + 2 i \mathrm{Im} (\alphag \alphae^*)\right) dt\right)\\
     = & 2 \chi \int_0^T \mathrm{Im} (\alphag \alphae^*) dt. 
\end{align*}
Comparing the first and last line with Eqs.~\eqref{eq:snr-opt} and \eqref{eq:def-gammam}, respectively, this equality shows indeed that the SNR,
when defined with optimal integration weights, and the measurement-induced dephasing
$\gammam$ are related by Eq.~(2), independent of the resonator parameters $\kappa$, $\chi$, and the functional form $\eps f(t)$ of the drive.

\section{Depletion tuneup}
\label{sec:cost}
\begin{figure}[ht!]   
\centering
\includegraphics{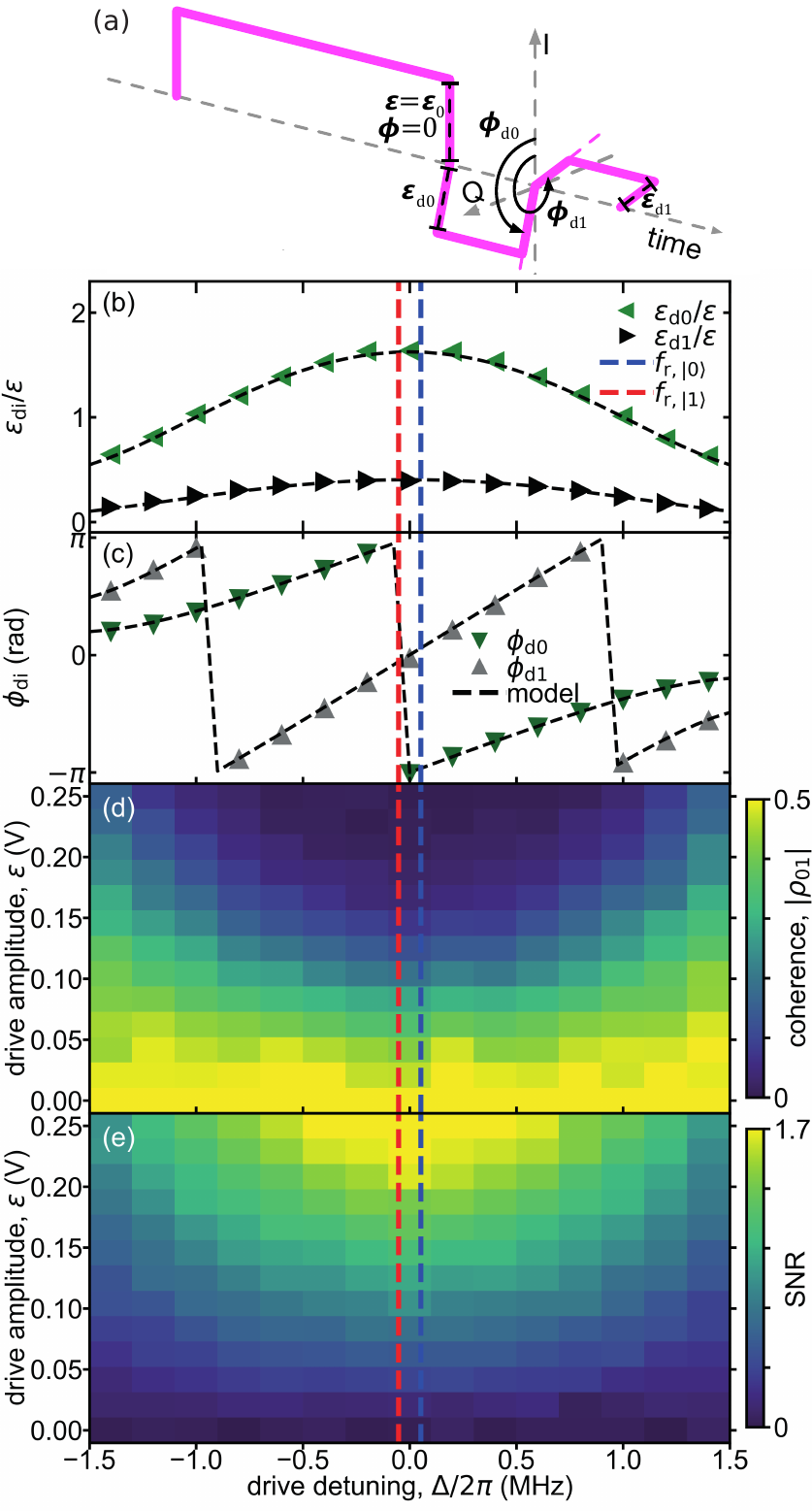}
\caption{\label{fig:detuning_supplement}   
	Depletion pulse parameters, coherence
	and SNR as a function of detuning. (a) The measurement pulse consists of a ramp-up of duration
  $\tauup=600~\ns$, fixed phase $\phi=0$ and amplitude $\eps$ (fixed during
  tuneup to $\eps=\epszero=0.25$ V) and two $200~\ns$ depletion segments
  ($\taud=400~\ns$) with each a tunable phase ($\phidzero$, $\phidone$) and
  amplitude ($\epsdzero$, $\epsdone$). (b,c) Depletion pulse parameters from the
	depletion optimizations used in Fig.~2. Dashed vertical lines
	indicate $f_\mathrm{r,\ket{0}}$ (blue) and $f_\mathrm{r,\ket{1}}$ (red). Dashed
	black curves are extracted from the linear model (see Sec.~\ref{sec:cost}). Coherence (d)
	and SNR (e) as a function of drive amplitude and detuning. At non-zero $\eps$,
	SNR is maximal (coherence is minimal) at the midpoint frequency
	$\Delta=0$ and decreases (increases) with detuning.   } 
\end{figure}

\begin{figure}[ht!]        
\centering          
\includegraphics{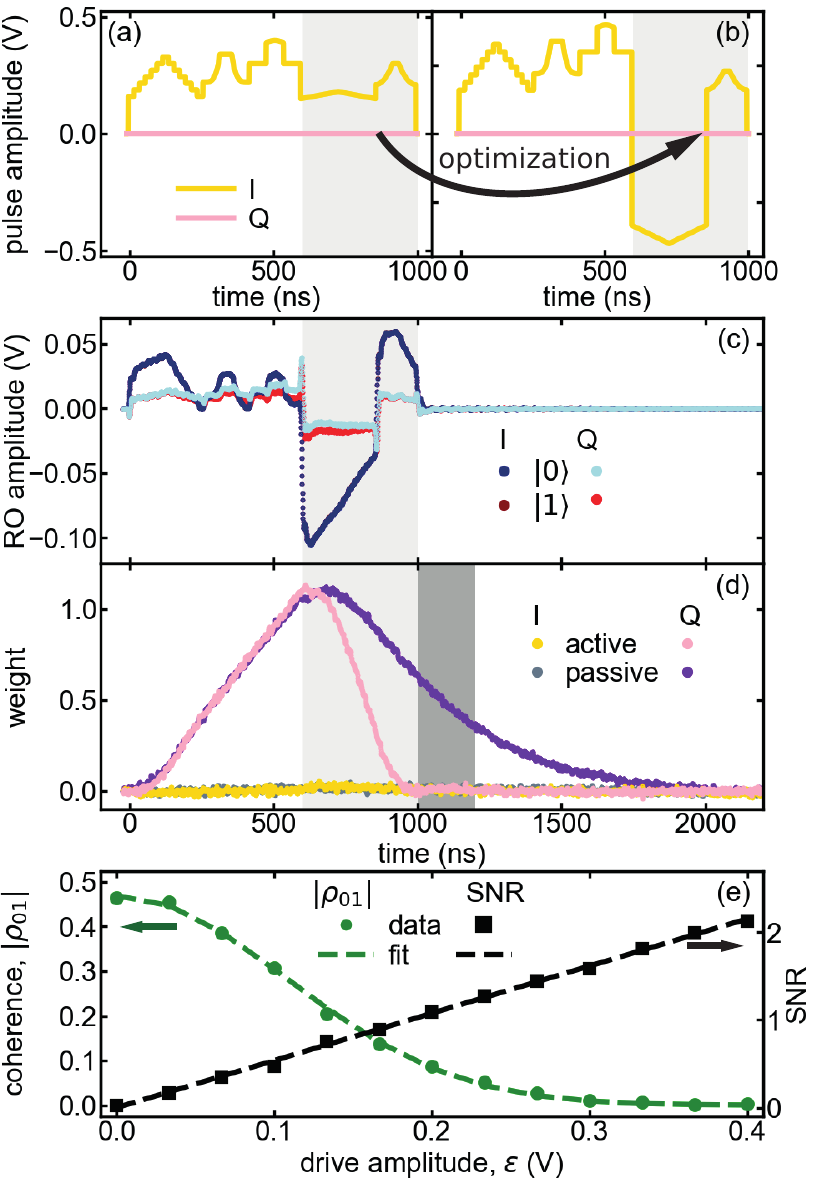}
\caption{\label{fig:amsterdam}   
	The three-step method for quantum efficiency
	extraction with a pulse envelope consisting of seventeenth-century Dutch canal
	house fa\c{c}ade outlines.   (a) Pulse envelope with five fa\c{c}ades, of which the first
	three ramp up the resonator with duration $\tauup=600~\ns$, fixed phase
	$\varphi=0$ and amplitude $\eps$ (fixed during tuneup to $\eps=\epszero=0.4$
	V) and the last two are $240~\ns$ and $160~\ns$ depletion segments
	($\taud=400~\ns$) with each a         tunable phase and amplitude. (b)
	Optimized depletion pulse with $\epsdzero=1.68\eps$,
	$\epsdone=0.58\eps$, $\phidzero=1.005 \pi$ rad, $\phidone=0.007 \pi$
	rad. (c) Averaged feedline transmission of the optimized         depletion
	pulse. The qubit is prepared in $\ket{0}$ (blue) and in
	$\ket{1}$ (red). (d) Optimal weight functions extracted for
	the depletion pulse (purple) and as a reference, weight functions are shown
	for passive depletion ($\epsdzero=\epsdone=0$ V). (d) Quantum
	efficiency extraction using 13 values of $\eps$. The best-fit values give $\etae=0.167\pm0.005$. } 
\end{figure}

\begin{figure}[ht!]   
\centering     
\includegraphics{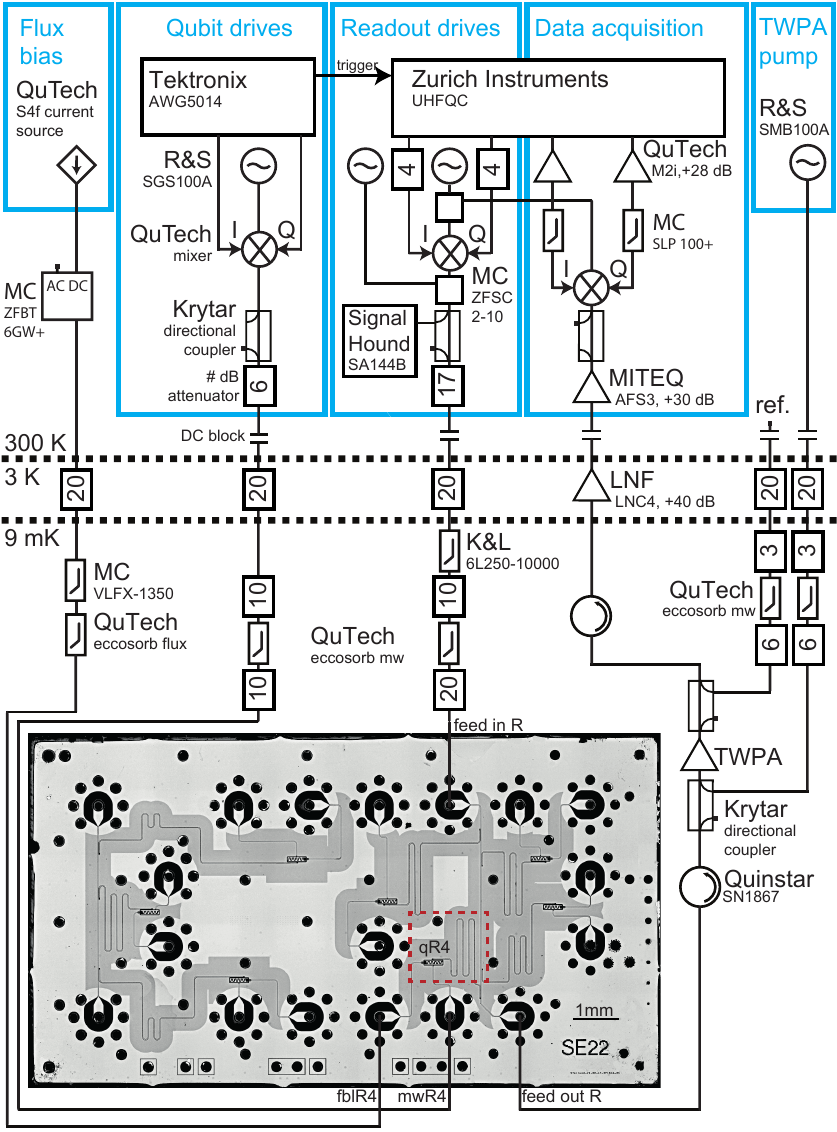}
\caption{\label{fig:setup}   
	Photograph of cQED chip (identical design as the one used) and
	complete wiring diagram of electronic components inside and outside the 
	$^3$He/$^4$He dilution refrigerator (Leiden Cryogenics CF-CS81). The test chip
	contains seven transmon qubits individually coupled to dedicated microwave drive
	lines, flux bias lines and readout resonators. The three (four)
	resonators on the left (right) side couple
	capacitively to the left (right) feedline traversing the chip from top to
	bottom. All 18 connections are made from the back side of the chip and reach the
	front through vertical coax lines~\cite{Versluis17e}. Each vertical coax line consists of a central
	through-silicon via (TSV) that carries the signal and seven surrounding TSVs
	acting as shield connecting the front and back side ground planes. Other,
	individual TSVs interconnect front side and back side ground planes to eliminate chip modes. } 
\end{figure}

Here, we provide details on the depletion tuneup. The depletion is tuned by optimizing the amplitude and phase of both depletion steps (Fig.~\ref{fig:detuning_supplement}) using the Nelder-Mead algorithm with a cost function that penalizes non-zero averaged transients for both $\ket{0}$ and $\ket{1}$ during a $\taucost=200~\ns$ time window after the depletion. The transients are obtained by preparing the qubit in $\ket{0}$ ($\ket{1}$) and averaging
the time-domain homodyne voltages $V_{\mathrm{I}, \ket{0}}$ and $V_{\mathrm{Q}, \ket{0}}$ ($V_{\mathrm{I}, \ket{1}}$ and $V_{\mathrm{Q}, \ket{1}}$) of the transmitted measurement pulse for $2^{15}$ repetitions. The cost function consists of four different terms. The first two null the transients in both quadratures post-depletion. The last two additionally penalize the difference between the transients for $\ket{0}$ and $\ket{1}$ with a tunable factor $d$. In the experiment, we found reliable convergence of the depletion tuneup for $d=10$.

\begin{align*} 
cost &= \sqrt{\int_{\tauup+\taud}^{\tauup+\taud+\taucost}\langle{V_{\mathrm{I},\ket{0}}}(t)\rangle^2+\langle{V_{\mathrm{Q},\ket{0}}}(t)\rangle^2 dt}\\ 
&+ \sqrt{\int_{\tauup+\taud}^{\tauup+\taud+\taucost}\langle {V_{\mathrm{I},\ket{1}}}(t)\rangle^2+\langle{V_{\mathrm{Q},\ket{1}}}(t)\rangle^2 dt}\\
&+d \sqrt{\int_{\tauup+\taud}^{\tauup+\taud+\taucost} \langle {V_{\mathrm{I},\ket{1}}}(t)-{V_{\mathrm{I},\ket{0}}}(t)\rangle^2 dt}\\
&+ d \sqrt{\int_{\tauup+\taud}^{\tauup+\taud+\taucost}\langle {V_{\mathrm{Q},\ket{1}}}(t)-{V_{\mathrm{Q},\ket{0}}}(t)\rangle^2 dt}.
\end{align*} 

In Figures~\ref{fig:detuning_supplement}(b,c), we show the obtained depletion pulse
parameters for different values of $\Delta$. As a comparison, we show the
parameters that are predicted by numerically integrating Eq.~(1), with resonator parameters extracted from Fig.~2(a), and numerically finding the depletion pulse parameters that lead to
$\alpha_{\ket{0, 1}}(T) = 0$.

\begin{thebibliography}{32}%
\makeatletter
\providecommand \@ifxundefined [1]{%
 \@ifx{#1\undefined}
}%
\providecommand \@ifnum [1]{%
 \ifnum #1\expandafter \@firstoftwo
 \else \expandafter \@secondoftwo
 \fi
}%
\providecommand \@ifx [1]{%
 \ifx #1\expandafter \@firstoftwo
 \else \expandafter \@secondoftwo
 \fi
}%
\providecommand \natexlab [1]{#1}%
\providecommand \enquote  [1]{``#1''}%
\providecommand \bibnamefont  [1]{#1}%
\providecommand \bibfnamefont [1]{#1}%
\providecommand \citenamefont [1]{#1}%
\providecommand \href@noop [0]{\@secondoftwo}%
\providecommand \href [0]{\begingroup \@sanitize@url \@href}%
\providecommand \@href[1]{\@@startlink{#1}\@@href}%
\providecommand \@@href[1]{\endgroup#1\@@endlink}%
\providecommand \@sanitize@url [0]{\catcode `\\12\catcode `\$12\catcode
  `\&12\catcode `\#12\catcode `\^12\catcode `\_12\catcode `\%12\relax}%
\providecommand \@@startlink[1]{}%
\providecommand \@@endlink[0]{}%
\providecommand \url  [0]{\begingroup\@sanitize@url \@url }%
\providecommand \@url [1]{\endgroup\@href {#1}{\urlprefix }}%
\providecommand \urlprefix  [0]{URL }%
\providecommand \Eprint [0]{\href }%
\providecommand \doibase [0]{http://dx.doi.org/}%
\providecommand \selectlanguage [0]{\@gobble}%
\providecommand \bibinfo  [0]{\@secondoftwo}%
\providecommand \bibfield  [0]{\@secondoftwo}%
\providecommand \translation [1]{[#1]}%
\providecommand \BibitemOpen [0]{}%
\providecommand \bibitemStop [0]{}%
\providecommand \bibitemNoStop [0]{.\EOS\space}%
\providecommand \EOS [0]{\spacefactor3000\relax}%
\providecommand \BibitemShut  [1]{\csname bibitem#1\endcsname}%
\let\auto@bib@innerbib\@empty
%</preamble>
\bibitem [{\citenamefont {DiVincenzo}(2009)}]{Divincenzo09}%
  \BibitemOpen
  \bibfield  {author} {\bibinfo {author} {\bibfnamefont {D.~P.}\ \bibnamefont
  {DiVincenzo}},\ }\href
  {https://arxiv.org/ct?url=http%3A%2F%2Fdx.doi.org%2F10%252E1088%2F0031-8949%2F2009%2FT137%2F014020&v=5eb633b4}
  {\bibfield  {journal} {\bibinfo  {journal} {Physica Scripta}\ }\textbf
  {\bibinfo {volume} {2009}},\ \bibinfo {pages} {014020} (\bibinfo {year}
  {2009})}\BibitemShut {NoStop}%
\bibitem [{\citenamefont {Terhal}(2015)}]{Terhal15}%
  \BibitemOpen
  \bibfield  {author} {\bibinfo {author} {\bibfnamefont {B.~M.}\ \bibnamefont
  {Terhal}},\ }\href {\doibase 10.1103/RevModPhys.87.307} {\bibfield  {journal}
  {\bibinfo  {journal} {Rev. Mod. Phys.}\ }\textbf {\bibinfo {volume} {87}},\
  \bibinfo {pages} {307} (\bibinfo {year} {2015})}\BibitemShut {NoStop}%
\bibitem [{\citenamefont {Blais}\ \emph {et~al.}(2004)\citenamefont {Blais},
  \citenamefont {Huang}, \citenamefont {Wallraff}, \citenamefont {Girvin},\
  and\ \citenamefont {Schoelkopf}}]{Blais04}%
  \BibitemOpen
  \bibfield  {author} {\bibinfo {author} {\bibfnamefont {A.}~\bibnamefont
  {Blais}}, \bibinfo {author} {\bibfnamefont {R.-S.}\ \bibnamefont {Huang}},
  \bibinfo {author} {\bibfnamefont {A.}~\bibnamefont {Wallraff}}, \bibinfo
  {author} {\bibfnamefont {S.~M.}\ \bibnamefont {Girvin}}, \ and\ \bibinfo
  {author} {\bibfnamefont {R.~J.}\ \bibnamefont {Schoelkopf}},\ }\href
  {https://link.aps.org/doi/10.1103/PhysRevA.69.062320} {\bibfield  {journal}
  {\bibinfo  {journal} {Phys. Rev. A}\ }\textbf {\bibinfo {volume} {69}},\
  \bibinfo {pages} {062320} (\bibinfo {year} {2004})}\BibitemShut {NoStop}%
\bibitem [{\citenamefont {Wallraff}\ \emph {et~al.}(2004)\citenamefont
  {Wallraff}, \citenamefont {Schuster}, \citenamefont {Blais}, \citenamefont
  {Frunzio}, \citenamefont {Huang}, \citenamefont {Majer}, \citenamefont
  {Kumar}, \citenamefont {Girvin},\ and\ \citenamefont
  {Schoelkopf}}]{Wallraff04}%
  \BibitemOpen
  \bibfield  {author} {\bibinfo {author} {\bibfnamefont {A.}~\bibnamefont
  {Wallraff}}, \bibinfo {author} {\bibfnamefont {D.~I.}\ \bibnamefont
  {Schuster}}, \bibinfo {author} {\bibfnamefont {A.}~\bibnamefont {Blais}},
  \bibinfo {author} {\bibfnamefont {L.}~\bibnamefont {Frunzio}}, \bibinfo
  {author} {\bibfnamefont {R.-S.}\ \bibnamefont {Huang}}, \bibinfo {author}
  {\bibfnamefont {J.}~\bibnamefont {Majer}}, \bibinfo {author} {\bibfnamefont
  {S.}~\bibnamefont {Kumar}}, \bibinfo {author} {\bibfnamefont {S.~M.}\
  \bibnamefont {Girvin}}, \ and\ \bibinfo {author} {\bibfnamefont {R.~J.}\
  \bibnamefont {Schoelkopf}},\ }\href
  {http://www.nature.com/nature/journal/v431/n7005/abs/nature02851.html}
  {\bibfield  {journal} {\bibinfo  {journal} {Nature}\ }\textbf {\bibinfo
  {volume} {431}},\ \bibinfo {pages} {162} (\bibinfo {year}
  {2004})}\BibitemShut {NoStop}%
\bibitem [{\citenamefont {Koch}\ \emph {et~al.}(2007)\citenamefont {Koch},
  \citenamefont {Yu}, \citenamefont {Gambetta}, \citenamefont {Houck},
  \citenamefont {Schuster}, \citenamefont {Majer}, \citenamefont {Blais},
  \citenamefont {Devoret}, \citenamefont {Girvin},\ and\ \citenamefont
  {Schoelkopf}}]{Koch07}%
  \BibitemOpen
  \bibfield  {author} {\bibinfo {author} {\bibfnamefont {J.}~\bibnamefont
  {Koch}}, \bibinfo {author} {\bibfnamefont {T.~M.}\ \bibnamefont {Yu}},
  \bibinfo {author} {\bibfnamefont {J.}~\bibnamefont {Gambetta}}, \bibinfo
  {author} {\bibfnamefont {A.~A.}\ \bibnamefont {Houck}}, \bibinfo {author}
  {\bibfnamefont {D.~I.}\ \bibnamefont {Schuster}}, \bibinfo {author}
  {\bibfnamefont {J.}~\bibnamefont {Majer}}, \bibinfo {author} {\bibfnamefont
  {A.}~\bibnamefont {Blais}}, \bibinfo {author} {\bibfnamefont {M.~H.}\
  \bibnamefont {Devoret}}, \bibinfo {author} {\bibfnamefont {S.~M.}\
  \bibnamefont {Girvin}}, \ and\ \bibinfo {author} {\bibfnamefont {R.~J.}\
  \bibnamefont {Schoelkopf}},\ }\href
  {http://journals.aps.org/pra/abstract/10.1103/PhysRevA.76.042319} {\bibfield
  {journal} {\bibinfo  {journal} {Phys. Rev. A}\ }\textbf {\bibinfo {volume}
  {76}},\ \bibinfo {pages} {042319} (\bibinfo {year} {2007})}\BibitemShut
  {NoStop}%
\bibitem [{\citenamefont {Clerk}\ \emph {et~al.}(2010)\citenamefont {Clerk},
  \citenamefont {Devoret}, \citenamefont {Girvin}, \citenamefont {Marquardt},\
  and\ \citenamefont {Schoelkopf}}]{Clerk10}%
  \BibitemOpen
  \bibfield  {author} {\bibinfo {author} {\bibfnamefont {A.~A.}\ \bibnamefont
  {Clerk}}, \bibinfo {author} {\bibfnamefont {M.~H.}\ \bibnamefont {Devoret}},
  \bibinfo {author} {\bibfnamefont {S.~M.}\ \bibnamefont {Girvin}}, \bibinfo
  {author} {\bibfnamefont {F.}~\bibnamefont {Marquardt}}, \ and\ \bibinfo
  {author} {\bibfnamefont {R.~J.}\ \bibnamefont {Schoelkopf}},\ }\href
  {\doibase 10.1103/RevModPhys.82.1155} {\bibfield  {journal} {\bibinfo
  {journal} {Reviews of Modern Physics}\ }\textbf {\bibinfo {volume} {82}},\
  \bibinfo {pages} {1155} (\bibinfo {year} {2010})},\ \Eprint
  {http://arxiv.org/abs/0810.4729} {0810.4729} \BibitemShut {NoStop}%
\bibitem [{\citenamefont {Castellanos-Beltran}\ \emph
  {et~al.}(2008)\citenamefont {Castellanos-Beltran}, \citenamefont {Irwin},
  \citenamefont {Hilton}, \citenamefont {Vale},\ and\ \citenamefont
  {Lehnert}}]{Castellanos-Beltran08}%
  \BibitemOpen
  \bibfield  {author} {\bibinfo {author} {\bibfnamefont {M.~A.}\ \bibnamefont
  {Castellanos-Beltran}}, \bibinfo {author} {\bibfnamefont {K.~D.}\
  \bibnamefont {Irwin}}, \bibinfo {author} {\bibfnamefont {G.~C.}\ \bibnamefont
  {Hilton}}, \bibinfo {author} {\bibfnamefont {L.~R.}\ \bibnamefont {Vale}}, \
  and\ \bibinfo {author} {\bibfnamefont {K.~W.}\ \bibnamefont {Lehnert}},\
  }\href {\doibase 10.1038/nphys1090} {\bibfield  {journal} {\bibinfo
  {journal} {Nat.\ Phys.}\ }\textbf {\bibinfo {volume} {4}},\ \bibinfo {pages}
  {929} (\bibinfo {year} {2008})}\BibitemShut {NoStop}%
\bibitem [{\citenamefont {Vijay}, \citenamefont {Devoret},\ and\ \citenamefont
  {Siddiqi}(2009)}]{Vijay09}%
  \BibitemOpen
  \bibfield  {author} {\bibinfo {author} {\bibfnamefont {R.}~\bibnamefont
  {Vijay}}, \bibinfo {author} {\bibfnamefont {M.~H.}\ \bibnamefont {Devoret}},
  \ and\ \bibinfo {author} {\bibfnamefont {I.}~\bibnamefont {Siddiqi}},\ }\href
  {http://aip.scitation.org/doi/10.1063/1.3224703} {\bibfield  {journal}
  {\bibinfo  {journal} {Rev. Sci. Instrum.}\ }\textbf {\bibinfo {volume}
  {80}},\ \bibinfo {pages} {111101} (\bibinfo {year} {2009})}\BibitemShut
  {NoStop}%
\bibitem [{\citenamefont {Bergeal}\ \emph {et~al.}(2010)\citenamefont
  {Bergeal}, \citenamefont {Schackert}, \citenamefont {Metcalfe}, \citenamefont
  {Vijay}, \citenamefont {Manucharyan}, \citenamefont {Frunzio}, \citenamefont
  {Prober}, \citenamefont {Schoelkopf}, \citenamefont {Girvin},\ and\
  \citenamefont {Devoret}}]{Bergeal10}%
  \BibitemOpen
  \bibfield  {author} {\bibinfo {author} {\bibfnamefont {N.}~\bibnamefont
  {Bergeal}}, \bibinfo {author} {\bibfnamefont {F.}~\bibnamefont {Schackert}},
  \bibinfo {author} {\bibfnamefont {M.}~\bibnamefont {Metcalfe}}, \bibinfo
  {author} {\bibfnamefont {R.}~\bibnamefont {Vijay}}, \bibinfo {author}
  {\bibfnamefont {V.~E.}\ \bibnamefont {Manucharyan}}, \bibinfo {author}
  {\bibfnamefont {L.}~\bibnamefont {Frunzio}}, \bibinfo {author} {\bibfnamefont
  {D.~E.}\ \bibnamefont {Prober}}, \bibinfo {author} {\bibfnamefont {R.~J.}\
  \bibnamefont {Schoelkopf}}, \bibinfo {author} {\bibfnamefont {S.~M.}\
  \bibnamefont {Girvin}}, \ and\ \bibinfo {author} {\bibfnamefont {M.~H.}\
  \bibnamefont {Devoret}},\ }\href {\doibase 10.1038/nature09035} {\bibfield
  {journal} {\bibinfo  {journal} {Nature}\ }\textbf {\bibinfo {volume} {465}},\
  \bibinfo {pages} {64} (\bibinfo {year} {2010})}\BibitemShut {NoStop}%
\bibitem [{\citenamefont {Mutus}\ \emph {et~al.}(2014)\citenamefont {Mutus},
  \citenamefont {White}, \citenamefont {Barends}, \citenamefont {Chen},
  \citenamefont {Chen}, \citenamefont {Chiaro}, \citenamefont {Dunsworth},
  \citenamefont {Jeffrey}, \citenamefont {Kelly}, \citenamefont {Megrant},
  \citenamefont {Neill}, \citenamefont {O'Malley}, \citenamefont {Roushan},
  \citenamefont {Sank}, \citenamefont {Vainsencher}, \citenamefont {Wenner},
  \citenamefont {Sundqvist}, \citenamefont {Cleland},\ and\ \citenamefont
  {Martinis}}]{Mutus14}%
  \BibitemOpen
  \bibfield  {author} {\bibinfo {author} {\bibfnamefont {J.~Y.}\ \bibnamefont
  {Mutus}}, \bibinfo {author} {\bibfnamefont {T.~C.}\ \bibnamefont {White}},
  \bibinfo {author} {\bibfnamefont {R.}~\bibnamefont {Barends}}, \bibinfo
  {author} {\bibfnamefont {Y.}~\bibnamefont {Chen}}, \bibinfo {author}
  {\bibfnamefont {Z.}~\bibnamefont {Chen}}, \bibinfo {author} {\bibfnamefont
  {B.}~\bibnamefont {Chiaro}}, \bibinfo {author} {\bibfnamefont
  {A.}~\bibnamefont {Dunsworth}}, \bibinfo {author} {\bibfnamefont
  {E.}~\bibnamefont {Jeffrey}}, \bibinfo {author} {\bibfnamefont
  {J.}~\bibnamefont {Kelly}}, \bibinfo {author} {\bibfnamefont
  {A.}~\bibnamefont {Megrant}}, \bibinfo {author} {\bibfnamefont
  {C.}~\bibnamefont {Neill}}, \bibinfo {author} {\bibfnamefont {P.~J.~J.}\
  \bibnamefont {O'Malley}}, \bibinfo {author} {\bibfnamefont {P.}~\bibnamefont
  {Roushan}}, \bibinfo {author} {\bibfnamefont {D.}~\bibnamefont {Sank}},
  \bibinfo {author} {\bibfnamefont {A.}~\bibnamefont {Vainsencher}}, \bibinfo
  {author} {\bibfnamefont {J.}~\bibnamefont {Wenner}}, \bibinfo {author}
  {\bibfnamefont {K.~M.}\ \bibnamefont {Sundqvist}}, \bibinfo {author}
  {\bibfnamefont {A.~N.}\ \bibnamefont {Cleland}}, \ and\ \bibinfo {author}
  {\bibfnamefont {J.~M.}\ \bibnamefont {Martinis}},\ }\href
  {http://scitation.aip.org/content/aip/journal/apl/104/26/10.1063/1.4886408}
  {\bibfield  {journal} {\bibinfo  {journal} {Appl. Phys. Lett.}\ }\textbf
  {\bibinfo {volume} {104}},\ \bibinfo {pages} {263513} (\bibinfo {year}
  {2014})}\BibitemShut {NoStop}%
\bibitem [{\citenamefont {Eichler}\ \emph {et~al.}(2014)\citenamefont
  {Eichler}, \citenamefont {Salathe}, \citenamefont {Mlynek}, \citenamefont
  {Schmidt},\ and\ \citenamefont {Wallraff}}]{Eichler14}%
  \BibitemOpen
  \bibfield  {author} {\bibinfo {author} {\bibfnamefont {C.}~\bibnamefont
  {Eichler}}, \bibinfo {author} {\bibfnamefont {Y.}~\bibnamefont {Salathe}},
  \bibinfo {author} {\bibfnamefont {J.}~\bibnamefont {Mlynek}}, \bibinfo
  {author} {\bibfnamefont {S.}~\bibnamefont {Schmidt}}, \ and\ \bibinfo
  {author} {\bibfnamefont {A.}~\bibnamefont {Wallraff}},\ }\href {\doibase
  10.1103/PhysRevLett.113.110502} {\bibfield  {journal} {\bibinfo  {journal}
  {Phys. Rev. Lett.}\ }\textbf {\bibinfo {volume} {113}},\ \bibinfo {pages}
  {110502} (\bibinfo {year} {2014})}\BibitemShut {NoStop}%
\bibitem [{\citenamefont {Johnson}\ \emph {et~al.}(2012)\citenamefont
  {Johnson}, \citenamefont {Macklin}, \citenamefont {Slichter}, \citenamefont
  {Vijay}, \citenamefont {Weingarten}, \citenamefont {Clarke},\ and\
  \citenamefont {Siddiqi}}]{Johnson12}%
  \BibitemOpen
  \bibfield  {author} {\bibinfo {author} {\bibfnamefont {J.~E.}\ \bibnamefont
  {Johnson}}, \bibinfo {author} {\bibfnamefont {C.}~\bibnamefont {Macklin}},
  \bibinfo {author} {\bibfnamefont {D.~H.}\ \bibnamefont {Slichter}}, \bibinfo
  {author} {\bibfnamefont {R.}~\bibnamefont {Vijay}}, \bibinfo {author}
  {\bibfnamefont {E.~B.}\ \bibnamefont {Weingarten}}, \bibinfo {author}
  {\bibfnamefont {J.}~\bibnamefont {Clarke}}, \ and\ \bibinfo {author}
  {\bibfnamefont {I.}~\bibnamefont {Siddiqi}},\ }\href {\doibase
  10.1103/PhysRevLett.109.050506} {\bibfield  {journal} {\bibinfo  {journal}
  {Phys. Rev. Lett.}\ }\textbf {\bibinfo {volume} {109}},\ \bibinfo {pages}
  {050506} (\bibinfo {year} {2012})}\BibitemShut {NoStop}%
\bibitem [{\citenamefont {Rist\`e}\ \emph {et~al.}(2012)\citenamefont
  {Rist\`e}, \citenamefont {van Leeuwen}, \citenamefont {Ku}, \citenamefont
  {Lehnert},\ and\ \citenamefont {DiCarlo}}]{Riste12}%
  \BibitemOpen
  \bibfield  {author} {\bibinfo {author} {\bibfnamefont {D.}~\bibnamefont
  {Rist\`e}}, \bibinfo {author} {\bibfnamefont {J.~G.}\ \bibnamefont {van
  Leeuwen}}, \bibinfo {author} {\bibfnamefont {H.-S.}\ \bibnamefont {Ku}},
  \bibinfo {author} {\bibfnamefont {K.~W.}\ \bibnamefont {Lehnert}}, \ and\
  \bibinfo {author} {\bibfnamefont {L.}~\bibnamefont {DiCarlo}},\ }\href
  {\doibase 10.1103/PhysRevLett.109.050507} {\bibfield  {journal} {\bibinfo
  {journal} {Phys. Rev. Lett.}\ }\textbf {\bibinfo {volume} {109}},\ \bibinfo
  {pages} {050507} (\bibinfo {year} {2012})}\BibitemShut {NoStop}%
\bibitem [{\citenamefont {Macklin}\ \emph {et~al.}(2015)\citenamefont
  {Macklin}, \citenamefont {O{\textquoteright}Brien}, \citenamefont {Hover},
  \citenamefont {Schwartz}, \citenamefont {Bolkhovsky}, \citenamefont {Zhang},
  \citenamefont {Oliver},\ and\ \citenamefont {Siddiqi}}]{Macklin15}%
  \BibitemOpen
  \bibfield  {author} {\bibinfo {author} {\bibfnamefont {C.}~\bibnamefont
  {Macklin}}, \bibinfo {author} {\bibfnamefont {K.}~\bibnamefont
  {O{\textquoteright}Brien}}, \bibinfo {author} {\bibfnamefont
  {D.}~\bibnamefont {Hover}}, \bibinfo {author} {\bibfnamefont {M.~E.}\
  \bibnamefont {Schwartz}}, \bibinfo {author} {\bibfnamefont {V.}~\bibnamefont
  {Bolkhovsky}}, \bibinfo {author} {\bibfnamefont {X.}~\bibnamefont {Zhang}},
  \bibinfo {author} {\bibfnamefont {W.~D.}\ \bibnamefont {Oliver}}, \ and\
  \bibinfo {author} {\bibfnamefont {I.}~\bibnamefont {Siddiqi}},\ }\href
  {http://www.sciencemag.org/cgi/doi/10.1126/science.aaa8525} {\bibfield
  {journal} {\bibinfo  {journal} {Science}\ }\textbf {\bibinfo {volume}
  {350}},\ \bibinfo {pages} {307} (\bibinfo {year} {2015})}\BibitemShut
  {NoStop}%
\bibitem [{\citenamefont {Vissers}\ \emph {et~al.}(2016)\citenamefont
  {Vissers}, \citenamefont {Erickson}, \citenamefont {Ku}, \citenamefont
  {Vale}, \citenamefont {Wu}, \citenamefont {Hilton},\ and\ \citenamefont
  {Pappas}}]{Vissers16}%
  \BibitemOpen
  \bibfield  {author} {\bibinfo {author} {\bibfnamefont {M.~R.}\ \bibnamefont
  {Vissers}}, \bibinfo {author} {\bibfnamefont {R.~P.}\ \bibnamefont
  {Erickson}}, \bibinfo {author} {\bibfnamefont {H.~S.}\ \bibnamefont {Ku}},
  \bibinfo {author} {\bibfnamefont {L.}~\bibnamefont {Vale}}, \bibinfo {author}
  {\bibfnamefont {X.}~\bibnamefont {Wu}}, \bibinfo {author} {\bibfnamefont
  {G.~C.}\ \bibnamefont {Hilton}}, \ and\ \bibinfo {author} {\bibfnamefont
  {D.~P.}\ \bibnamefont {Pappas}},\ }\href
  {http://aip.scitation.org/doi/10.1063/1.4937922} {\bibfield  {journal}
  {\bibinfo  {journal} {Appl. Phys. Lett.}\ }\textbf {\bibinfo {volume}
  {108}},\ \bibinfo {pages} {012601} (\bibinfo {year} {2016})}\BibitemShut
  {NoStop}%
\bibitem [{\citenamefont {Gambetta}\ \emph {et~al.}(2006)\citenamefont
  {Gambetta}, \citenamefont {Blais}, \citenamefont {Schuster}, \citenamefont
  {Wallraff}, \citenamefont {Frunzio}, \citenamefont {Majer}, \citenamefont
  {Devoret}, \citenamefont {Girvin},\ and\ \citenamefont
  {Schoelkopf}}]{Gambetta06}%
  \BibitemOpen
  \bibfield  {author} {\bibinfo {author} {\bibfnamefont {J.}~\bibnamefont
  {Gambetta}}, \bibinfo {author} {\bibfnamefont {A.}~\bibnamefont {Blais}},
  \bibinfo {author} {\bibfnamefont {D.~I.}\ \bibnamefont {Schuster}}, \bibinfo
  {author} {\bibfnamefont {A.}~\bibnamefont {Wallraff}}, \bibinfo {author}
  {\bibfnamefont {L.}~\bibnamefont {Frunzio}}, \bibinfo {author} {\bibfnamefont
  {J.}~\bibnamefont {Majer}}, \bibinfo {author} {\bibfnamefont {M.~H.}\
  \bibnamefont {Devoret}}, \bibinfo {author} {\bibfnamefont {S.~M.}\
  \bibnamefont {Girvin}}, \ and\ \bibinfo {author} {\bibfnamefont {R.~J.}\
  \bibnamefont {Schoelkopf}},\ }\href {\doibase 10.1103/PhysRevA.74.042318}
  {\bibfield  {journal} {\bibinfo  {journal} {Phys. Rev. A}\ }\textbf {\bibinfo
  {volume} {74}},\ \bibinfo {pages} {042318} (\bibinfo {year}
  {2006})}\BibitemShut {NoStop}%
\bibitem [{Foo()}]{FootnoteBultink17}%
  \BibitemOpen
  \href@noop {} {}\bibinfo {howpublished} {This definition of quantum efficiency
  applies to phase-preserving amplification where the unavoidable quantum noise
  from the idler mode of the amplifier is included in the quantum limit. Using
  this definition, $\eta=1$ corresponds to an ideal phase preserving
  amplification. Furthermore, we define the $\SNR$ as the average 
  separation of the integrated homodyne voltage histograms as obtained from single-shot readout experiments
  for $\ket{0}$ and $\ket{1}$, divided by their average standard deviation (see supplementary material). }\BibitemShut {Stop}%
\bibitem [{\citenamefont {Vijay}, \citenamefont {Slichter},\ and\ \citenamefont
  {Siddiqi}(2011)}]{Vijay11}%
  \BibitemOpen
  \bibfield  {author} {\bibinfo {author} {\bibfnamefont {R.}~\bibnamefont
  {Vijay}}, \bibinfo {author} {\bibfnamefont {D.~H.}\ \bibnamefont {Slichter}},
  \ and\ \bibinfo {author} {\bibfnamefont {I.}~\bibnamefont {Siddiqi}},\ }\href
  {https://journals.aps.org/prl/abstract/10.1103/PhysRevLett.106.110502}
  {\bibfield  {journal} {\bibinfo  {journal} {Phys. Rev. Lett.}\ }\textbf
  {\bibinfo {volume} {106}},\ \bibinfo {pages} {110502} (\bibinfo {year}
  {2011})}\BibitemShut {NoStop}%
\bibitem [{\citenamefont {Hatridge}\ \emph {et~al.}(2013)\citenamefont
  {Hatridge}, \citenamefont {Shankar}, \citenamefont {Mirrahimi}, \citenamefont
  {Schackert}, \citenamefont {Geerlings}, \citenamefont {Brecht}, \citenamefont
  {Sliwa}, \citenamefont {Abdo}, \citenamefont {Frunzio}, \citenamefont
  {Girvin}, \citenamefont {Schoelkopf},\ and\ \citenamefont
  {Devoret}}]{Hatridge13}%
  \BibitemOpen
  \bibfield  {author} {\bibinfo {author} {\bibfnamefont {M.}~\bibnamefont
  {Hatridge}}, \bibinfo {author} {\bibfnamefont {S.}~\bibnamefont {Shankar}},
  \bibinfo {author} {\bibfnamefont {M.}~\bibnamefont {Mirrahimi}}, \bibinfo
  {author} {\bibfnamefont {F.}~\bibnamefont {Schackert}}, \bibinfo {author}
  {\bibfnamefont {K.}~\bibnamefont {Geerlings}}, \bibinfo {author}
  {\bibfnamefont {T.}~\bibnamefont {Brecht}}, \bibinfo {author} {\bibfnamefont
  {K.}~\bibnamefont {Sliwa}}, \bibinfo {author} {\bibfnamefont
  {B.}~\bibnamefont {Abdo}}, \bibinfo {author} {\bibfnamefont {L.}~\bibnamefont
  {Frunzio}}, \bibinfo {author} {\bibfnamefont {S.}~\bibnamefont {Girvin}},
  \bibinfo {author} {\bibfnamefont {R.}~\bibnamefont {Schoelkopf}}, \ and\
  \bibinfo {author} {\bibfnamefont {M.}~\bibnamefont {Devoret}},\ }\href
  {http://science.sciencemag.org/content/339/6116/178} {\bibfield  {journal}
  {\bibinfo  {journal} {Science}\ }\textbf {\bibinfo {volume} {339}},\ \bibinfo
  {pages} {178} (\bibinfo {year} {2013})}\BibitemShut {NoStop}%
\bibitem [{\citenamefont {Jeffrey}\ \emph {et~al.}(2014)\citenamefont
  {Jeffrey}, \citenamefont {Sank}, \citenamefont {Mutus}, \citenamefont
  {White}, \citenamefont {Kelly}, \citenamefont {Barends}, \citenamefont
  {Chen}, \citenamefont {Chen}, \citenamefont {Chiaro}, \citenamefont
  {Dunsworth}, \citenamefont {Megrant}, \citenamefont {O'Malley}, \citenamefont
  {Neill}, \citenamefont {Roushan}, \citenamefont {Vainsencher}, \citenamefont
  {Wenner}, \citenamefont {Cleland},\ and\ \citenamefont
  {Martinis}}]{Jeffrey14}%
  \BibitemOpen
  \bibfield  {author} {\bibinfo {author} {\bibfnamefont {E.}~\bibnamefont
  {Jeffrey}}, \bibinfo {author} {\bibfnamefont {D.}~\bibnamefont {Sank}},
  \bibinfo {author} {\bibfnamefont {J.~Y.}\ \bibnamefont {Mutus}}, \bibinfo
  {author} {\bibfnamefont {T.~C.}\ \bibnamefont {White}}, \bibinfo {author}
  {\bibfnamefont {J.}~\bibnamefont {Kelly}}, \bibinfo {author} {\bibfnamefont
  {R.}~\bibnamefont {Barends}}, \bibinfo {author} {\bibfnamefont
  {Y.}~\bibnamefont {Chen}}, \bibinfo {author} {\bibfnamefont {Z.}~\bibnamefont
  {Chen}}, \bibinfo {author} {\bibfnamefont {B.}~\bibnamefont {Chiaro}},
  \bibinfo {author} {\bibfnamefont {A.}~\bibnamefont {Dunsworth}}, \bibinfo
  {author} {\bibfnamefont {A.}~\bibnamefont {Megrant}}, \bibinfo {author}
  {\bibfnamefont {P.~J.~J.}\ \bibnamefont {O'Malley}}, \bibinfo {author}
  {\bibfnamefont {C.}~\bibnamefont {Neill}}, \bibinfo {author} {\bibfnamefont
  {P.}~\bibnamefont {Roushan}}, \bibinfo {author} {\bibfnamefont
  {A.}~\bibnamefont {Vainsencher}}, \bibinfo {author} {\bibfnamefont
  {J.}~\bibnamefont {Wenner}}, \bibinfo {author} {\bibfnamefont {A.~N.}\
  \bibnamefont {Cleland}}, \ and\ \bibinfo {author} {\bibfnamefont {J.~M.}\
  \bibnamefont {Martinis}},\ }\href
  {https://journals.aps.org/prl/abstract/10.1103/PhysRevLett.112.190504}
  {\bibfield  {journal} {\bibinfo  {journal} {Phys. Rev. Lett.}\ }\textbf
  {\bibinfo {volume} {112}},\ \bibinfo {pages} {190504} (\bibinfo {year}
  {2014})}\BibitemShut {NoStop}%
\bibitem [{\citenamefont {Liu}\ \emph {et~al.}(2017)\citenamefont {Liu},
  \citenamefont {Li}, \citenamefont {Dai}, \citenamefont {Zhang}, \citenamefont
  {Xue}, \citenamefont {Tan}, \citenamefont {Yu},\ and\ \citenamefont
  {Yu}}]{Liu17}%
  \BibitemOpen
  \bibfield  {author} {\bibinfo {author} {\bibfnamefont {Q.}~\bibnamefont
  {Liu}}, \bibinfo {author} {\bibfnamefont {M.}~\bibnamefont {Li}}, \bibinfo
  {author} {\bibfnamefont {K.}~\bibnamefont {Dai}}, \bibinfo {author}
  {\bibfnamefont {K.}~\bibnamefont {Zhang}}, \bibinfo {author} {\bibfnamefont
  {G.}~\bibnamefont {Xue}}, \bibinfo {author} {\bibfnamefont {X.}~\bibnamefont
  {Tan}}, \bibinfo {author} {\bibfnamefont {H.}~\bibnamefont {Yu}}, \ and\
  \bibinfo {author} {\bibfnamefont {Y.}~\bibnamefont {Yu}},\ }\href {\doibase
  10.1063/1.4985435} {\bibfield  {journal} {\bibinfo  {journal} {Applied
  Physics Letters}\ }\textbf {\bibinfo {volume} {110}},\ \bibinfo {pages}
  {232602} (\bibinfo {year} {2017})}\BibitemShut {NoStop}%
\bibitem [{\citenamefont {Reagor}\ \emph {et~al.}(2017)\citenamefont {Reagor},
  \citenamefont {Osborn}, \citenamefont {Tezak}, \citenamefont {Staley},
  \citenamefont {Prawiroatmodjo}, \citenamefont {Scheer}, \citenamefont
  {Alidoust}, \citenamefont {Sete}, \citenamefont {Didier}, \citenamefont
  {da~Silva}, \citenamefont {Acala}, \citenamefont {Angeles}, \citenamefont
  {Bestwick}, \citenamefont {Block}, \citenamefont {Bloom}, \citenamefont
  {Bradley}, \citenamefont {Bui}, \citenamefont {Caldwell}, \citenamefont
  {Capelluto}, \citenamefont {Chilcott}, \citenamefont {Cordova}, \citenamefont
  {Crossman}, \citenamefont {Curtis}, \citenamefont {Deshpande}, \citenamefont
  {Bouayadi}, \citenamefont {Girshovich}, \citenamefont {Hong}, \citenamefont
  {Hudson}, \citenamefont {Karalekas}, \citenamefont {Kuang}, \citenamefont
  {Lenihan}, \citenamefont {Manenti}, \citenamefont {Manning}, \citenamefont
  {Marshall}, \citenamefont {Mohan}, \citenamefont {O'Brien}, \citenamefont
  {Otterbach}, \citenamefont {Papageorge}, \citenamefont {Paquette},
  \citenamefont {Pelstring}, \citenamefont {Polloreno}, \citenamefont {Rawat},
  \citenamefont {Ryan}, \citenamefont {Renzas}, \citenamefont {Rubin},
  \citenamefont {Russell}, \citenamefont {Rust}, \citenamefont {Scarabelli},
  \citenamefont {Selvanayagam}, \citenamefont {Sinclair}, \citenamefont
  {Smith}, \citenamefont {Suska}, \citenamefont {To}, \citenamefont
  {Vahidpour}, \citenamefont {Vodrahalli}, \citenamefont {Whyland},
  \citenamefont {Yadav}, \citenamefont {Zeng},\ and\ \citenamefont
  {Rigetti}}]{Reagor17}%
  \BibitemOpen
  \bibfield  {author} {\bibinfo {author} {\bibfnamefont {M.}~\bibnamefont
  {Reagor}}, \bibinfo {author} {\bibfnamefont {C.~B.}\ \bibnamefont {Osborn}},
  \bibinfo {author} {\bibfnamefont {N.}~\bibnamefont {Tezak}}, \bibinfo
  {author} {\bibfnamefont {A.}~\bibnamefont {Staley}}, \bibinfo {author}
  {\bibfnamefont {G.}~\bibnamefont {Prawiroatmodjo}}, \bibinfo {author}
  {\bibfnamefont {M.}~\bibnamefont {Scheer}}, \bibinfo {author} {\bibfnamefont
  {N.}~\bibnamefont {Alidoust}}, \bibinfo {author} {\bibfnamefont {E.~A.}\
  \bibnamefont {Sete}}, \bibinfo {author} {\bibfnamefont {N.}~\bibnamefont
  {Didier}}, \bibinfo {author} {\bibfnamefont {M.~P.}\ \bibnamefont
  {da~Silva}}, \bibinfo {author} {\bibfnamefont {E.}~\bibnamefont {Acala}},
  \bibinfo {author} {\bibfnamefont {J.}~\bibnamefont {Angeles}}, \bibinfo
  {author} {\bibfnamefont {A.}~\bibnamefont {Bestwick}}, \bibinfo {author}
  {\bibfnamefont {M.}~\bibnamefont {Block}}, \bibinfo {author} {\bibfnamefont
  {B.}~\bibnamefont {Bloom}}, \bibinfo {author} {\bibfnamefont
  {A.}~\bibnamefont {Bradley}}, \bibinfo {author} {\bibfnamefont
  {C.}~\bibnamefont {Bui}}, \bibinfo {author} {\bibfnamefont {S.}~\bibnamefont
  {Caldwell}}, \bibinfo {author} {\bibfnamefont {L.}~\bibnamefont {Capelluto}},
  \bibinfo {author} {\bibfnamefont {R.}~\bibnamefont {Chilcott}}, \bibinfo
  {author} {\bibfnamefont {J.}~\bibnamefont {Cordova}}, \bibinfo {author}
  {\bibfnamefont {G.}~\bibnamefont {Crossman}}, \bibinfo {author}
  {\bibfnamefont {M.}~\bibnamefont {Curtis}}, \bibinfo {author} {\bibfnamefont
  {S.}~\bibnamefont {Deshpande}}, \bibinfo {author} {\bibfnamefont {T.~E.}\
  \bibnamefont {Bouayadi}}, \bibinfo {author} {\bibfnamefont {D.}~\bibnamefont
  {Girshovich}}, \bibinfo {author} {\bibfnamefont {S.}~\bibnamefont {Hong}},
  \bibinfo {author} {\bibfnamefont {A.}~\bibnamefont {Hudson}}, \bibinfo
  {author} {\bibfnamefont {P.}~\bibnamefont {Karalekas}}, \bibinfo {author}
  {\bibfnamefont {K.}~\bibnamefont {Kuang}}, \bibinfo {author} {\bibfnamefont
  {M.}~\bibnamefont {Lenihan}}, \bibinfo {author} {\bibfnamefont
  {R.}~\bibnamefont {Manenti}}, \bibinfo {author} {\bibfnamefont
  {T.}~\bibnamefont {Manning}}, \bibinfo {author} {\bibfnamefont
  {J.}~\bibnamefont {Marshall}}, \bibinfo {author} {\bibfnamefont
  {Y.}~\bibnamefont {Mohan}}, \bibinfo {author} {\bibfnamefont
  {W.}~\bibnamefont {O'Brien}}, \bibinfo {author} {\bibfnamefont
  {J.}~\bibnamefont {Otterbach}}, \bibinfo {author} {\bibfnamefont
  {A.}~\bibnamefont {Papageorge}}, \bibinfo {author} {\bibfnamefont {J.~P.}\
  \bibnamefont {Paquette}}, \bibinfo {author} {\bibfnamefont {M.}~\bibnamefont
  {Pelstring}}, \bibinfo {author} {\bibfnamefont {A.}~\bibnamefont
  {Polloreno}}, \bibinfo {author} {\bibfnamefont {V.}~\bibnamefont {Rawat}},
  \bibinfo {author} {\bibfnamefont {C.~A.}\ \bibnamefont {Ryan}}, \bibinfo
  {author} {\bibfnamefont {R.}~\bibnamefont {Renzas}}, \bibinfo {author}
  {\bibfnamefont {N.}~\bibnamefont {Rubin}}, \bibinfo {author} {\bibfnamefont
  {D.}~\bibnamefont {Russell}}, \bibinfo {author} {\bibfnamefont
  {M.}~\bibnamefont {Rust}}, \bibinfo {author} {\bibfnamefont {D.}~\bibnamefont
  {Scarabelli}}, \bibinfo {author} {\bibfnamefont {M.}~\bibnamefont
  {Selvanayagam}}, \bibinfo {author} {\bibfnamefont {R.}~\bibnamefont
  {Sinclair}}, \bibinfo {author} {\bibfnamefont {R.}~\bibnamefont {Smith}},
  \bibinfo {author} {\bibfnamefont {M.}~\bibnamefont {Suska}}, \bibinfo
  {author} {\bibfnamefont {T.~W.}\ \bibnamefont {To}}, \bibinfo {author}
  {\bibfnamefont {M.}~\bibnamefont {Vahidpour}}, \bibinfo {author}
  {\bibfnamefont {N.}~\bibnamefont {Vodrahalli}}, \bibinfo {author}
  {\bibfnamefont {T.}~\bibnamefont {Whyland}}, \bibinfo {author} {\bibfnamefont
  {K.}~\bibnamefont {Yadav}}, \bibinfo {author} {\bibfnamefont
  {W.}~\bibnamefont {Zeng}}, \ and\ \bibinfo {author} {\bibfnamefont {C.~T.}\
  \bibnamefont {Rigetti}},\ }\href {https://arxiv.org/abs/1706.06570}
  {\bibfield  {journal} {\bibinfo  {journal} {arXiv:1706.06570}\ } (\bibinfo
  {year} {2017})}\BibitemShut {NoStop}%
\bibitem [{\citenamefont {Takita}\ \emph {et~al.}(2017)\citenamefont {Takita},
  \citenamefont {Cross}, \citenamefont {C\'orcoles}, \citenamefont {Chow},\
  and\ \citenamefont {Gambetta}}]{Takita17}%
  \BibitemOpen
  \bibfield  {author} {\bibinfo {author} {\bibfnamefont {M.}~\bibnamefont
  {Takita}}, \bibinfo {author} {\bibfnamefont {A.~W.}\ \bibnamefont {Cross}},
  \bibinfo {author} {\bibfnamefont {A.~D.}\ \bibnamefont {C\'orcoles}},
  \bibinfo {author} {\bibfnamefont {J.~M.}\ \bibnamefont {Chow}}, \ and\
  \bibinfo {author} {\bibfnamefont {J.~M.}\ \bibnamefont {Gambetta}},\ }\href
  {https://journals.aps.org/prl/abstract/10.1103/PhysRevLett.119.180501}
  {\bibfield  {journal} {\bibinfo  {journal} {Phys. Rev. Lett.}\ }\textbf
  {\bibinfo {volume} {119}},\ \bibinfo {pages} {180501} (\bibinfo {year}
  {2017})}\BibitemShut {NoStop}%
\bibitem [{\citenamefont {Neill}\ \emph {et~al.}(2017)\citenamefont {Neill},
  \citenamefont {Roushan}, \citenamefont {Kechedzhi}, \citenamefont {Boixo},
  \citenamefont {Isakov}, \citenamefont {Smelyanskiy}, \citenamefont {Barends},
  \citenamefont {Burkett}, \citenamefont {Chen}, \citenamefont {Chen},
  \citenamefont {Chiaro}, \citenamefont {Dunsworth}, \citenamefont {Fowler},
  \citenamefont {Foxen}, \citenamefont {Graff}, \citenamefont {Jeffrey},
  \citenamefont {Kelly}, \citenamefont {Lucero}, \citenamefont {Megrant},
  \citenamefont {Mutus}, \citenamefont {Neeley}, \citenamefont {Quintana},
  \citenamefont {Sank}, \citenamefont {Vainsencher}, \citenamefont {Wenner},
  \citenamefont {White}, \citenamefont {Neven},\ and\ \citenamefont
  {Martinis}}]{Neill17}%
  \BibitemOpen
  \bibfield  {author} {\bibinfo {author} {\bibfnamefont {C.}~\bibnamefont
  {Neill}}, \bibinfo {author} {\bibfnamefont {P.}~\bibnamefont {Roushan}},
  \bibinfo {author} {\bibfnamefont {K.}~\bibnamefont {Kechedzhi}}, \bibinfo
  {author} {\bibfnamefont {S.}~\bibnamefont {Boixo}}, \bibinfo {author}
  {\bibfnamefont {S.~V.}\ \bibnamefont {Isakov}}, \bibinfo {author}
  {\bibfnamefont {V.}~\bibnamefont {Smelyanskiy}}, \bibinfo {author}
  {\bibfnamefont {R.}~\bibnamefont {Barends}}, \bibinfo {author} {\bibfnamefont
  {B.}~\bibnamefont {Burkett}}, \bibinfo {author} {\bibfnamefont
  {Y.}~\bibnamefont {Chen}}, \bibinfo {author} {\bibfnamefont {Z.}~\bibnamefont
  {Chen}}, \bibinfo {author} {\bibfnamefont {B.}~\bibnamefont {Chiaro}},
  \bibinfo {author} {\bibfnamefont {A.}~\bibnamefont {Dunsworth}}, \bibinfo
  {author} {\bibfnamefont {A.}~\bibnamefont {Fowler}}, \bibinfo {author}
  {\bibfnamefont {B.}~\bibnamefont {Foxen}}, \bibinfo {author} {\bibfnamefont
  {R.}~\bibnamefont {Graff}}, \bibinfo {author} {\bibfnamefont
  {E.}~\bibnamefont {Jeffrey}}, \bibinfo {author} {\bibfnamefont
  {J.}~\bibnamefont {Kelly}}, \bibinfo {author} {\bibfnamefont
  {E.}~\bibnamefont {Lucero}}, \bibinfo {author} {\bibfnamefont
  {A.}~\bibnamefont {Megrant}}, \bibinfo {author} {\bibfnamefont
  {J.}~\bibnamefont {Mutus}}, \bibinfo {author} {\bibfnamefont
  {M.}~\bibnamefont {Neeley}}, \bibinfo {author} {\bibfnamefont
  {C.}~\bibnamefont {Quintana}}, \bibinfo {author} {\bibfnamefont
  {D.}~\bibnamefont {Sank}}, \bibinfo {author} {\bibfnamefont {A.}~\bibnamefont
  {Vainsencher}}, \bibinfo {author} {\bibfnamefont {J.}~\bibnamefont {Wenner}},
  \bibinfo {author} {\bibfnamefont {T.~C.}\ \bibnamefont {White}}, \bibinfo
  {author} {\bibfnamefont {H.}~\bibnamefont {Neven}}, \ and\ \bibinfo {author}
  {\bibfnamefont {J.~M.}\ \bibnamefont {Martinis}},\ }\href
  {https://arxiv.org/abs/1709.06678} {\bibfield  {journal} {\bibinfo  {journal}
  {arXiv:1709.06678}\ } (\bibinfo {year} {2017})}\BibitemShut {NoStop}%
\bibitem [{\citenamefont {Versluis}\ \emph {et~al.}(2017)\citenamefont
  {Versluis}, \citenamefont {Poletto}, \citenamefont {Khammassi}, \citenamefont
  {Tarasinski}, \citenamefont {Haider}, \citenamefont {Michalak}, \citenamefont
  {Bruno}, \citenamefont {Bertels},\ and\ \citenamefont
  {DiCarlo}}]{Versluis17}%
  \BibitemOpen
  \bibfield  {author} {\bibinfo {author} {\bibfnamefont {R.}~\bibnamefont
  {Versluis}}, \bibinfo {author} {\bibfnamefont {S.}~\bibnamefont {Poletto}},
  \bibinfo {author} {\bibfnamefont {N.}~\bibnamefont {Khammassi}}, \bibinfo
  {author} {\bibfnamefont {B.}~\bibnamefont {Tarasinski}}, \bibinfo {author}
  {\bibfnamefont {N.}~\bibnamefont {Haider}}, \bibinfo {author} {\bibfnamefont
  {D.~J.}\ \bibnamefont {Michalak}}, \bibinfo {author} {\bibfnamefont
  {A.}~\bibnamefont {Bruno}}, \bibinfo {author} {\bibfnamefont
  {K.}~\bibnamefont {Bertels}}, \ and\ \bibinfo {author} {\bibfnamefont
  {L.}~\bibnamefont {DiCarlo}},\ }\href {\doibase
  10.1103/PhysRevApplied.8.034021} {\bibfield  {journal} {\bibinfo  {journal}
  {Phys. Rev. Applied}\ }\textbf {\bibinfo {volume} {8}},\ \bibinfo {pages}
  {034021} (\bibinfo {year} {2017})}\BibitemShut {NoStop}%
\bibitem [{\citenamefont {McClure}\ \emph {et~al.}(2016)\citenamefont
  {McClure}, \citenamefont {Paik}, \citenamefont {Bishop}, \citenamefont
  {Steffen}, \citenamefont {Chow},\ and\ \citenamefont {Gambetta}}]{McClure16}%
  \BibitemOpen
  \bibfield  {author} {\bibinfo {author} {\bibfnamefont {D.~T.}\ \bibnamefont
  {McClure}}, \bibinfo {author} {\bibfnamefont {H.}~\bibnamefont {Paik}},
  \bibinfo {author} {\bibfnamefont {L.~S.}\ \bibnamefont {Bishop}}, \bibinfo
  {author} {\bibfnamefont {M.}~\bibnamefont {Steffen}}, \bibinfo {author}
  {\bibfnamefont {J.~M.}\ \bibnamefont {Chow}}, \ and\ \bibinfo {author}
  {\bibfnamefont {J.~M.}\ \bibnamefont {Gambetta}},\ }\href
  {https://journals.aps.org/prapplied/abstract/10.1103/PhysRevApplied.5.011001}
  {\bibfield  {journal} {\bibinfo  {journal} {Phys. Rev. Appl.}\ }\textbf
  {\bibinfo {volume} {5}},\ \bibinfo {pages} {011001} (\bibinfo {year}
  {2016})}\BibitemShut {NoStop}%
\bibitem [{\citenamefont {Bultink}\ \emph {et~al.}(2016)\citenamefont
  {Bultink}, \citenamefont {Rol}, \citenamefont {O'Brien}, \citenamefont {Fu},
  \citenamefont {Dikken}, \citenamefont {Dickel}, \citenamefont {Vermeulen},
  \citenamefont {de~Sterke}, \citenamefont {Bruno}, \citenamefont {Schouten},\
  and\ \citenamefont {DiCarlo}}]{Bultink16}%
  \BibitemOpen
  \bibfield  {author} {\bibinfo {author} {\bibfnamefont {C.~C.}\ \bibnamefont
  {Bultink}}, \bibinfo {author} {\bibfnamefont {M.~A.}\ \bibnamefont {Rol}},
  \bibinfo {author} {\bibfnamefont {T.~E.}\ \bibnamefont {O'Brien}}, \bibinfo
  {author} {\bibfnamefont {X.}~\bibnamefont {Fu}}, \bibinfo {author}
  {\bibfnamefont {B.~C.~S.}\ \bibnamefont {Dikken}}, \bibinfo {author}
  {\bibfnamefont {C.}~\bibnamefont {Dickel}}, \bibinfo {author} {\bibfnamefont
  {R.~F.~L.}\ \bibnamefont {Vermeulen}}, \bibinfo {author} {\bibfnamefont
  {J.~C.}\ \bibnamefont {de~Sterke}}, \bibinfo {author} {\bibfnamefont
  {A.}~\bibnamefont {Bruno}}, \bibinfo {author} {\bibfnamefont {R.~N.}\
  \bibnamefont {Schouten}}, \ and\ \bibinfo {author} {\bibfnamefont
  {L.}~\bibnamefont {DiCarlo}},\ }\href
  {https://link.aps.org/doi/10.1103/PhysRevApplied.6.034008} {\bibfield
  {journal} {\bibinfo  {journal} {Phys. Rev. Appl.}\ }\textbf {\bibinfo
  {volume} {6}},\ \bibinfo {pages} {034008} (\bibinfo {year}
  {2016})}\BibitemShut {NoStop}%
\bibitem [{\citenamefont {Ryan}\ \emph {et~al.}(2015)\citenamefont {Ryan},
  \citenamefont {Johnson}, \citenamefont {Gambetta}, \citenamefont {Chow},
  \citenamefont {da~Silva}, \citenamefont {Dial},\ and\ \citenamefont
  {Ohki}}]{Ryan15}%
  \BibitemOpen
  \bibfield  {author} {\bibinfo {author} {\bibfnamefont {C.~A.}\ \bibnamefont
  {Ryan}}, \bibinfo {author} {\bibfnamefont {B.~R.}\ \bibnamefont {Johnson}},
  \bibinfo {author} {\bibfnamefont {J.~M.}\ \bibnamefont {Gambetta}}, \bibinfo
  {author} {\bibfnamefont {J.~M.}\ \bibnamefont {Chow}}, \bibinfo {author}
  {\bibfnamefont {M.~P.}\ \bibnamefont {da~Silva}}, \bibinfo {author}
  {\bibfnamefont {O.~E.}\ \bibnamefont {Dial}}, \ and\ \bibinfo {author}
  {\bibfnamefont {T.~A.}\ \bibnamefont {Ohki}},\ }\href
  {https://journals.aps.org/pra/pdf/10.1103/PhysRevA.91.022118} {\bibfield
  {journal} {\bibinfo  {journal} {Phys. Rev. A}\ }\textbf {\bibinfo {volume}
  {91}},\ \bibinfo {pages} {022118} (\bibinfo {year} {2015})}\BibitemShut
  {NoStop}%
\bibitem [{\citenamefont {Magesan}\ \emph {et~al.}(2015)\citenamefont
  {Magesan}, \citenamefont {Gambetta}, \citenamefont {C\'orcoles},\ and\
  \citenamefont {Chow}}]{Magesan15}%
  \BibitemOpen
  \bibfield  {author} {\bibinfo {author} {\bibfnamefont {E.}~\bibnamefont
  {Magesan}}, \bibinfo {author} {\bibfnamefont {J.~M.}\ \bibnamefont
  {Gambetta}}, \bibinfo {author} {\bibfnamefont {A.~D.}\ \bibnamefont
  {C\'orcoles}}, \ and\ \bibinfo {author} {\bibfnamefont {J.~M.}\ \bibnamefont
  {Chow}},\ }\href
  {https://journals.aps.org/prl/pdf/10.1103/PhysRevLett.114.200501} {\bibfield
  {journal} {\bibinfo  {journal} {Phys. Rev. Lett.}\ }\textbf {\bibinfo
  {volume} {114}},\ \bibinfo {pages} {200501} (\bibinfo {year}
  {2015})}\BibitemShut {NoStop}%
\bibitem [{\citenamefont {Frisk~Kockum}, \citenamefont {Tornberg},\ and\
  \citenamefont {Johansson}(2012)}]{FriskKockum12}%
  \BibitemOpen
  \bibfield  {author} {\bibinfo {author} {\bibfnamefont {A.}~\bibnamefont
  {Frisk~Kockum}}, \bibinfo {author} {\bibfnamefont {L.}~\bibnamefont
  {Tornberg}}, \ and\ \bibinfo {author} {\bibfnamefont {G.}~\bibnamefont
  {Johansson}},\ }\href {https://link.aps.org/doi/10.1103/PhysRevA.85.052318}
  {\bibfield  {journal} {\bibinfo  {journal} {Phys. Rev. A}\ }\textbf {\bibinfo
  {volume} {85}},\ \bibinfo {pages} {052318} (\bibinfo {year}
  {2012})}\BibitemShut {NoStop}%
\bibitem [{\citenamefont {Walter}\ \emph {et~al.}(2017)\citenamefont {Walter},
  \citenamefont {Kurpiers}, \citenamefont {Gasparinetti}, \citenamefont
  {Magnard}, \citenamefont {Poto{\v{c}}nik}, \citenamefont {Salath{\'{e}}},
  \citenamefont {Pechal}, \citenamefont {Mondal}, \citenamefont {Oppliger},
  \citenamefont {Eichler},\ and\ \citenamefont {Wallraff}}]{Walter17}%
  \BibitemOpen
  \bibfield  {author} {\bibinfo {author} {\bibfnamefont {T.}~\bibnamefont
  {Walter}}, \bibinfo {author} {\bibfnamefont {P.}~\bibnamefont {Kurpiers}},
  \bibinfo {author} {\bibfnamefont {S.}~\bibnamefont {Gasparinetti}}, \bibinfo
  {author} {\bibfnamefont {P.}~\bibnamefont {Magnard}}, \bibinfo {author}
  {\bibfnamefont {A.}~\bibnamefont {Poto{\v{c}}nik}}, \bibinfo {author}
  {\bibfnamefont {Y.}~\bibnamefont {Salath{\'{e}}}}, \bibinfo {author}
  {\bibfnamefont {M.}~\bibnamefont {Pechal}}, \bibinfo {author} {\bibfnamefont
  {M.}~\bibnamefont {Mondal}}, \bibinfo {author} {\bibfnamefont
  {M.}~\bibnamefont {Oppliger}}, \bibinfo {author} {\bibfnamefont
  {C.}~\bibnamefont {Eichler}}, \ and\ \bibinfo {author} {\bibfnamefont
  {A.}~\bibnamefont {Wallraff}},\ }\href
  {https://journals.aps.org/prapplied/abstract/10.1103/PhysRevApplied.7.054020}
  {\bibfield  {journal} {\bibinfo  {journal} {Phys. Rev. Appl.}\ }\textbf
  {\bibinfo {volume} {7}},\ \bibinfo {pages} {054020} (\bibinfo {year}
  {2017})}\BibitemShut {NoStop}%
\end{thebibliography}

\begin{thebibliography}{4}%
\makeatletter
\providecommand \@ifxundefined [1]{%
 \@ifx{#1\undefined}
}%
\providecommand \@ifnum [1]{%
 \ifnum #1\expandafter \@firstoftwo
 \else \expandafter \@secondoftwo
 \fi
}%
\providecommand \@ifx [1]{%
 \ifx #1\expandafter \@firstoftwo
 \else \expandafter \@secondoftwo
 \fi
}%
\providecommand \natexlab [1]{#1}%
\providecommand \enquote  [1]{``#1''}%
\providecommand \bibnamefont  [1]{#1}%
\providecommand \bibfnamefont [1]{#1}%
\providecommand \citenamefont [1]{#1}%
\providecommand \href@noop [0]{\@secondoftwo}%
\providecommand \href [0]{\begingroup \@sanitize@url \@href}%
\providecommand \@href[1]{\@@startlink{#1}\@@href}%
\providecommand \@@href[1]{\endgroup#1\@@endlink}%
\providecommand \@sanitize@url [0]{\catcode `\\12\catcode `\$12\catcode
  `\&12\catcode `\#12\catcode `\^12\catcode `\_12\catcode `\%12\relax}%
\providecommand \@@startlink[1]{}%
\providecommand \@@endlink[0]{}%
\providecommand \url  [0]{\begingroup\@sanitize@url \@url }%
\providecommand \@url [1]{\endgroup\@href {#1}{\urlprefix }}%
\providecommand \urlprefix  [0]{URL }%
\providecommand \Eprint [0]{\href }%
\providecommand \doibase [0]{http://dx.doi.org/}%
\providecommand \selectlanguage [0]{\@gobble}%
\providecommand \bibinfo  [0]{\@secondoftwo}%
\providecommand \bibfield  [0]{\@secondoftwo}%
\providecommand \translation [1]{[#1]}%
\providecommand \BibitemOpen [0]{}%
\providecommand \bibitemStop [0]{}%
\providecommand \bibitemNoStop [0]{.\EOS\space}%
\providecommand \EOS [0]{\spacefactor3000\relax}%
\providecommand \BibitemShut  [1]{\csname bibitem#1\endcsname}%
\let\auto@bib@innerbib\@empty
%</preamble>
\bibitem [{\citenamefont {Frisk~Kockum}, \citenamefont {Tornberg},\ and\
  \citenamefont {Johansson}(2012)}]{FriskKockum12e}%
  \BibitemOpen
  \bibfield  {author} {\bibinfo {author} {\bibfnamefont {A.}~\bibnamefont
  {Frisk~Kockum}}, \bibinfo {author} {\bibfnamefont {L.}~\bibnamefont
  {Tornberg}}, \ and\ \bibinfo {author} {\bibfnamefont {G.}~\bibnamefont
  {Johansson}},\ }\href {https://link.aps.org/doi/10.1103/PhysRevA.85.052318}
  {\bibfield  {journal} {\bibinfo  {journal} {Phys. Rev. A}\ }\textbf {\bibinfo
  {volume} {85}},\ \bibinfo {pages} {052318} (\bibinfo {year}
  {2012})}\BibitemShut {NoStop}%
\bibitem [{\citenamefont {Ryan}\ \emph {et~al.}(2015)\citenamefont {Ryan},
  \citenamefont {Johnson}, \citenamefont {Gambetta}, \citenamefont {Chow},
  \citenamefont {da~Silva}, \citenamefont {Dial},\ and\ \citenamefont
  {Ohki}}]{Ryan15e}%
  \BibitemOpen
  \bibfield  {author} {\bibinfo {author} {\bibfnamefont {C.~A.}\ \bibnamefont
  {Ryan}}, \bibinfo {author} {\bibfnamefont {B.~R.}\ \bibnamefont {Johnson}},
  \bibinfo {author} {\bibfnamefont {J.~M.}\ \bibnamefont {Gambetta}}, \bibinfo
  {author} {\bibfnamefont {J.~M.}\ \bibnamefont {Chow}}, \bibinfo {author}
  {\bibfnamefont {M.~P.}\ \bibnamefont {da~Silva}}, \bibinfo {author}
  {\bibfnamefont {O.~E.}\ \bibnamefont {Dial}}, \ and\ \bibinfo {author}
  {\bibfnamefont {T.~A.}\ \bibnamefont {Ohki}},\ }\href
  {https://journals.aps.org/pra/pdf/10.1103/PhysRevA.91.022118} {\bibfield
  {journal} {\bibinfo  {journal} {Phys. Rev. A}\ }\textbf {\bibinfo {volume}
  {91}},\ \bibinfo {pages} {022118} (\bibinfo {year} {2015})}\BibitemShut
  {NoStop}%
\bibitem [{\citenamefont {Magesan}\ \emph {et~al.}(2015)\citenamefont
  {Magesan}, \citenamefont {Gambetta}, \citenamefont {C\'orcoles},\ and\
  \citenamefont {Chow}}]{Magesan15e}%
  \BibitemOpen
  \bibfield  {author} {\bibinfo {author} {\bibfnamefont {E.}~\bibnamefont
  {Magesan}}, \bibinfo {author} {\bibfnamefont {J.~M.}\ \bibnamefont
  {Gambetta}}, \bibinfo {author} {\bibfnamefont {A.~D.}\ \bibnamefont
  {C\'orcoles}}, \ and\ \bibinfo {author} {\bibfnamefont {J.~M.}\ \bibnamefont
  {Chow}},\ }\href
  {https://journals.aps.org/prl/pdf/10.1103/PhysRevLett.114.200501} {\bibfield
  {journal} {\bibinfo  {journal} {Phys. Rev. Lett.}\ }\textbf {\bibinfo
  {volume} {114}},\ \bibinfo {pages} {200501} (\bibinfo {year}
  {2015})}\BibitemShut {NoStop}%
\bibitem [{\citenamefont {Versluis}\ \emph {et~al.}(2017)\citenamefont
  {Versluis}, \citenamefont {Poletto}, \citenamefont {Khammassi}, \citenamefont
  {Tarasinski}, \citenamefont {Haider}, \citenamefont {Michalak}, \citenamefont
  {Bruno}, \citenamefont {Bertels},\ and\ \citenamefont
  {DiCarlo}}]{Versluis17e}%
  \BibitemOpen
  \bibfield  {author} {\bibinfo {author} {\bibfnamefont {R.}~\bibnamefont
  {Versluis}}, \bibinfo {author} {\bibfnamefont {S.}~\bibnamefont {Poletto}},
  \bibinfo {author} {\bibfnamefont {N.}~\bibnamefont {Khammassi}}, \bibinfo
  {author} {\bibfnamefont {B.}~\bibnamefont {Tarasinski}}, \bibinfo {author}
  {\bibfnamefont {N.}~\bibnamefont {Haider}}, \bibinfo {author} {\bibfnamefont
  {D.~J.}\ \bibnamefont {Michalak}}, \bibinfo {author} {\bibfnamefont
  {A.}~\bibnamefont {Bruno}}, \bibinfo {author} {\bibfnamefont
  {K.}~\bibnamefont {Bertels}}, \ and\ \bibinfo {author} {\bibfnamefont
  {L.}~\bibnamefont {DiCarlo}},\ }\href {\doibase
  10.1103/PhysRevApplied.8.034021} {\bibfield  {journal} {\bibinfo  {journal}
  {Phys. Rev. Applied}\ }\textbf {\bibinfo {volume} {8}},\ \bibinfo {pages}
  {034021} (\bibinfo {year} {2017})}\BibitemShut {NoStop}%
\end{thebibliography}
\end{document}